%% file: acl_latex.tex
\PassOptionsToPackage{table}{xcolor}
\documentclass{lightspeed}

\definecolor{titlebg}{HTML}{EEF3FA}

\fancypagestyle{firstheader}{%
  \fancyhf{}
  \fancyhead[R]{%
    \includegraphics[height=1.02cm,trim={67 34 76 44},clip]{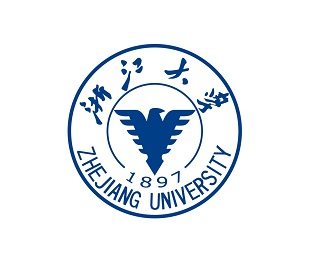}%
    \hspace{0.38cm}%
    \includegraphics[height=1.02cm,trim={143 142 142 137},clip]{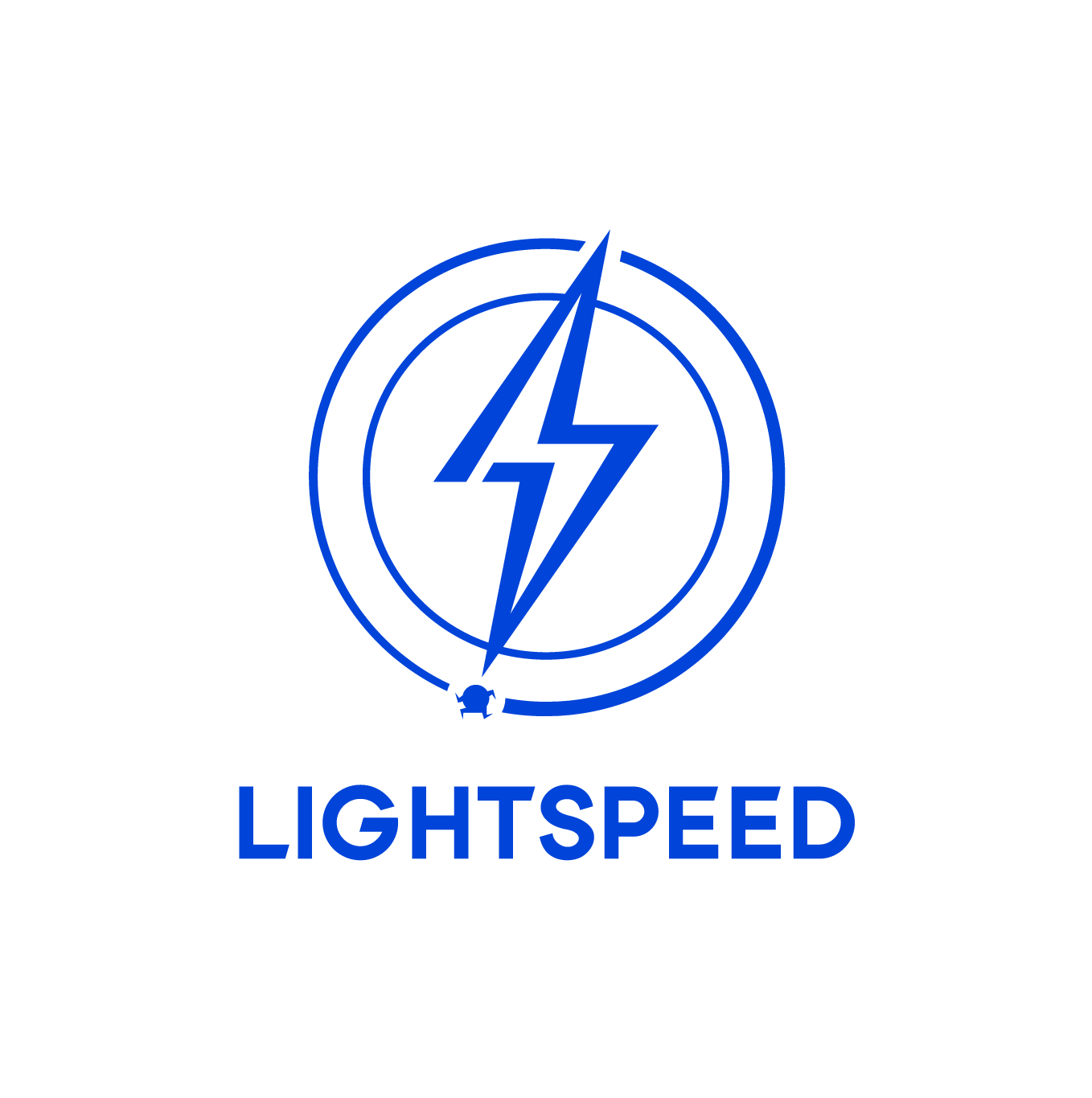}%
  }
  \renewcommand{\headrulewidth}{0.8pt}
  \renewcommand{\headrule}{\hbox to\headwidth{\color{metabg}\leaders\hrule height \headrulewidth\hfill}}
}

\renewcommand{\title}[1]{\newcommand{\titlelist}{{\LARGE\sffamily\bfseries #1}}}
\renewcommand\authorformat[2][]{{\sffamily\bfseries\color{metabg}#2$^{#1}$}}
\renewcommand\affiliationformat[2][]{{\small\color{metabg}$^{#1}$#2}}
\renewcommand\contributionformat[2][]{{\small\color{metabg}$^{#1}$#2}}
\renewcommand\emailformat[2][]{{\footnotesize\texttt{\href{mailto:#2}{#2}}}}
\renewcommand\metadataformat[2][]{{\small{\sffamily\bfseries #1:} #2}}

\renewcommand{\mymaketitle}{%
  \begingroup
  \setlength{\parindent}{0pt}
  \setlength{\parskip}{0pt}
  \vspace*{-0.35cm}
  \begin{tcolorbox}[enhanced,colback=titlebg,colframe=titlebg,
      boxrule=0pt,arc=9pt,left=0.55cm,right=0.55cm,
      top=0.45cm,bottom=0.40cm,before skip=0pt,after skip=0.45cm]
    {\raggedright\nohyphens
      {\setstretch{1.08}\titlelist\par}
      \vspace{0.28cm}
      {\authorlist\par}
      \vspace{0.14cm}
      {\affiliationlist\par}
      \ifdefempty{\contributionlist}{}{\vspace{0.06cm}\contributionlist\par}
      \ifdefempty{\emaillist}{}{\vspace{0.10cm}{\color{metabg}\emaillist\par}}
      \vspace{0.32cm}
      {\small\setstretch{1.08}\color{metafg}\abstractlist\par}
      \ifdefempty{\headercontentlist}{}{\vspace{0.26cm}\headercontentlist}
      \ifdefempty{\metadatalist}{}{\metadatalist\par}
    }
  \end{tcolorbox}
  \endgroup
  \FloatBarrier
}

\microtypesetup{expansion=false}
\setcitestyle{square,comma,numbers,sort&compress}

\usepackage[T1]{fontenc}
\usepackage[utf8]{inputenc}
\usepackage{microtype}
\usepackage{inconsolata}

\usepackage{graphicx}
\usepackage{wrapfig}
\usepackage{needspace}
\usepackage{placeins}
\usepackage{flafter}
\usepackage{booktabs}
\usepackage{multirow}
\usepackage{amsmath}
\usepackage{amssymb}
\usepackage{url}
\usepackage{enumitem}
\usepackage{xspace}
\usepackage{listings}

\colorlet{takeawaybg}{titlebg}
\newtcolorbox{takeawaybox}{
  enhanced,
  breakable,
  colback=takeawaybg,
  colframe=takeawaybg,
  boxrule=0pt,
  arc=0.8mm,
  left=1.2mm,
  right=1.2mm,
  top=0.9mm,
  bottom=0.9mm
}

\definecolor{decS}{HTML}{A6D5CB}
\definecolor{decM}{HTML}{C6E3DC}
\definecolor{decL}{HTML}{E2F0EC}
\definecolor{incL}{HTML}{F7E0CE}
\definecolor{incM}{HTML}{EDBB94}

\input{paper_macros}

\title{Training the Whole Agent, Not Just the Model}
\author[1,*,\dagger]{Lingxiao Tang}
\author[2,\dagger]{Wei Tao}
\author[3,\ddagger]{Yiwen Guo}
\author[1,\ddagger]{Lingfeng Bao}

\affiliation[1]{Zhejiang University}
\affiliation[2]{LIGHTSPEED}
\affiliation[3]{Independent Researcher}
\renewcommand{\affiliationlist}{%
  \affiliationformat[1]{Zhejiang University}\quad
  \affiliationformat[2]{LIGHTSPEED}\quad
  \affiliationformat[3]{Independent Researcher}%
}
\renewcommand{\emaillist}{%
  \emailformat{lingxiaotang@zju.edu.cn}\quad
  \emailformat{wtao@ieee.org}\\[2pt]
  \emailformat{guoyiwen89@gmail.com}\quad
  \emailformat{lingfengbao@zju.edu.cn}%
}
\abstract{\input{sections/abstract}}
\headercontent{%
  {\small{\sffamily\bfseries Code:} \texttt{\href{https://github.com/lingxiaotang/HarnessAdapt-RL}{github.com/lingxiaotang/HarnessAdapt-RL}}\par}%
}
\date{October 2026}
\hypersetup{pdftitle={Training the Whole Agent, Not Just the Model},
  pdfauthor={Lingxiao Tang, Wei Tao, Yiwen Guo, Lingfeng Bao}}

\begin{document}
\thispagestyle{firstheader}
\maketitle
\begingroup
\renewcommand{\thefootnote}{\fnsymbol{footnote}}
\footnotetext[1]{Work is done during internship at LIGHTSPEED.}
\footnotetext[2]{Equal contribution: Lingxiao Tang and Wei Tao.}
\footnotetext[3]{\raggedright Corresponding authors: Yiwen Guo (\href{mailto:guoyiwen89@gmail.com}{\nolinkurl{guoyiwen89@gmail.com}}) and Lingfeng Bao (\href{mailto:lingfengbao@zju.edu.cn}{\nolinkurl{lingfengbao@zju.edu.cn}}).}
\endgroup
\setcounter{footnote}{0}

\input{sections/introduction}
\input{sections/preliminaries}
\input{sections/method}
\input{sections/experimental_setup}
\input{sections/main_results}
\input{sections/analysis}

\input{sections/discussion_evolution}
\input{sections/related_work}
\input{sections/conclusion}
\input{sections/data_availability}
\input{appendix/limitations}

\FloatBarrier
\bibliography{custom}
\bibliographystyle{unsrtnat}

\clearpage
\beginappendix
\input{appendix/data_harness}
\FloatBarrier
\input{appendix/implementation}
\FloatBarrier
\input{appendix/additional_results}
\FloatBarrier
\input{appendix/harness_evolution}
\FloatBarrier
\input{appendix/prompts}

\end{document}

%% file: paper_macros.tex
\usepackage{array}
\usepackage{fancyvrb}
\newcolumntype{L}[1]{>{\raggedright\arraybackslash}p{#1}}
\newcommand{\tabgroup}[1]{\multicolumn{2}{@{}l}{\textit{#1}}\\}
\usepackage{listings}
\definecolor{promptframe}{HTML}{B8B8B8}
\definecolor{promptback}{HTML}{F2F2F2}
\definecolor{promptcodeback}{HTML}{FFFFFF}
\definecolor{promptcoderule}{HTML}{E6E6E6}
\lstdefinestyle{promptcode}{
  basicstyle=\ttfamily\fontsize{9}{10.5}\selectfont,
  columns=fullflexible,
  keepspaces=true,
  breaklines=true,
  breakatwhitespace=false,
  breakindent=1em,
  showstringspaces=false,
  upquote=true,
  backgroundcolor=\color{promptcodeback},
  frame=none,
  xleftmargin=0pt,
  xrightmargin=0pt,
  aboveskip=3pt,
  belowskip=0pt,
  literate={---}{{-{}-{}-}}3 {--}{{-{}-}}2 {–}{{\textendash}}1 {→}{{$\rightarrow$}}1
}
\newtcolorbox{promptbox}[1]{
  enhanced,
  colback=white, colframe=promptframe, colbacktitle=promptback,
  boxrule=0.5pt,
  arc=0pt, outer arc=0pt,
  fonttitle=\bfseries\small, coltitle=black,
  title={#1},
  left=7pt, right=7pt, top=5pt, bottom=5pt,
  before skip=0pt, after skip=0pt
}
\newcommand{\promptpart}[1]{\par\noindent{\bfseries\small #1}\par\vspace{1pt}}
\newcommand{\promptsep}{\par\vspace{5pt}\noindent{\color{promptframe}\rule{\linewidth}{0.4pt}}\par\vspace{4pt}}

\graphicspath{{figures/}}

\newcommand{\appname}{HarnessAdapt-RL\xspace}

\newsavebox{\sidepanelbox}
\makeatletter
\newcommand{\finishsidepanel}{%
  \par
  \ifnum\c@WF@wrappedlines>1
    \vspace{\dimexpr\c@WF@wrappedlines\baselineskip\relax}%
  \fi
  \WFclear
}
\makeatother
\newcommand{\sidepanel}[4][]{%
  \finishsidepanel
  \FloatBarrier
  \sbox{\sidepanelbox}{%
    \begin{minipage}{#3}
      \setlength{\columnwidth}{\linewidth}
      \centering
      \captionsetup{type=#2,justification=raggedright,singlelinecheck=false}
      #4
    \end{minipage}%
  }%
  \Needspace{\dimexpr\ht\sidepanelbox+\dp\sidepanelbox+2\intextsep+4\baselineskip\relax}%
  #1%
  \begin{wrapfloat}{#2}{r}{#3}
    \usebox{\sidepanelbox}%
  \end{wrapfloat}%
}
\newcommand{\flowpanel}[4][]{%
  \finishsidepanel
  \FloatBarrier
  \sbox{\sidepanelbox}{%
    \begin{minipage}{#3}
      \setlength{\columnwidth}{\linewidth}
      \centering
      \captionsetup{type=#2,justification=raggedright,singlelinecheck=false}
      #4
    \end{minipage}%
  }%
  #1%
  \begin{wrapfloat}{#2}{R}{#3}
    \usebox{\sidepanelbox}%
  \end{wrapfloat}%
}
\newenvironment{compacttable}[1][htbp]{%
  \begin{table}[#1]\centering
  \begin{minipage}{0.72\textwidth}\centering
  \captionsetup{justification=raggedright,singlelinecheck=true}
}{%
  \end{minipage}\end{table}
}

\makeatletter
\g@addto@macro\UrlBreaks{\do\a\do\b\do\c\do\d\do\e\do\f\do\g\do\h\do\i\do\j\do\k\do\l\do\m\do\n\do\o\do\p\do\q\do\r\do\s\do\t\do\u\do\v\do\w\do\x\do\y\do\z\do\A\do\B\do\C\do\D\do\E\do\F\do\G\do\H\do\I\do\J\do\K\do\L\do\M\do\N\do\O\do\P\do\Q\do\R\do\S\do\T\do\U\do\V\do\W\do\X\do\Y\do\Z\do\0\do\1\do\2\do\3\do\4\do\5\do\6\do\7\do\8\do\9}
\makeatother


%% file: sections/abstract.tex
Training coding agents typically updates the policy while keeping the harness fixed. During training, trajectories reveal flawed repair strategies and interaction problems that vary across tasks and change as the policy learns. We introduce \appname, which trains the policy and adapts its harness during reinforcement learning. For flawed repair strategies, a frozen teacher generates task-specific mechanism-level skills. For interaction problems recurring across tasks, a separate teacher model evolves tools and hooks, keeping each change only if it improves the observed pass rate or the observed trainable rate without lowering either. The policy trains with the adapted harness, and its new trajectories guide further adaptation. With a 9B base model, \appname improves pass@1 over GRPO by 9.8, 12.6, and 6.4 percentage points on SWE-bench Verified, Multilingual, and Pro, and outperforms GRPO with the human-designed OpenHands harness. Gains from mechanism-level skills persist after their removal at evaluation. Re-evolving tools and hooks as the policy changes outperforms freezing early versions. Removing them at deployment, in contrast, costs up to 7.7 percentage points. Training therefore produces a model--harness pair: the trained policy and the evolved tools and hooks, with skills removed at deployment.

%% file: sections/introduction.tex
\section{Introduction}
\label{sec:intro}

A coding agent operates as a model--harness pair: the model selects actions, while the harness provides guidance, tools, and runtime feedback. Agent training typically improves the policy while keeping the harness fixed~\citep{jain2025r2e,yang2025swesmith,xi2026agentgymrl}. The harness shapes the interaction trajectories used to update the policy, affecting both the effectiveness of training and the performance of the trained agent~\citep{chen2026coharness}. Prior work has explored both harness optimization and training-time guidance. Previous methods optimize the harness while keeping the policy fixed~\citep{lee2026metaharness,jiang2026harnessevolve,du2026livingharness}. Other methods use skills and training guidance to improve policy learning from trajectories~\citep{xiao2026socraticswe,hhd2026,pats2026}. Co-Harness alternates harness updates with supervised fine-tuning on trajectories generated by the improved harness~\citep{chen2026coharness}.

We study harness adaptation during reinforcement learning for coding agents. During training, trajectories expose flawed repair strategies, such as incorrect diagnoses or fixes that only address symptoms. They also reveal interaction problems, such as failed editing commands, excessive output, or insufficient verification. This evidence can guide changes to skills, tools, and runtime hooks that help alleviate these problems. As the problems vary across tasks and change with the policy, the harness needs to adapt throughout training (Appendix~\ref{app:verification}). We therefore ask how to design skills for policy learning, how to design and select tools and hooks as the policy changes, and which harness components to retain at deployment.

To address these questions, we introduce \textbf{\appname}, which trains the policy and adapts its harness during reinforcement learning. In our setting, skills are textual guidance supplied to the policy during training. First, we compare three skill-guidance variants and use mechanism-level skills that explain when and why a repair strategy applies without providing concrete answers. A frozen teacher model generates these task-specific skills from training trajectories. Second, a separate teacher model identifies recurring interaction and feedback problems across tasks and proposes changes to tools and hooks. Harness selection considers both task success and the availability of rollouts for policy updates (Section~\ref{sec:acceptance}). Each proposal is kept only if it improves the observed pass rate or the observed trainable rate without lowering either (Appendix~\ref{app:candidate-generation}). We refer to this process as tool and hook evolution. The policy then trains with GRPO~\citep{shao2024deepseekmath} on fresh rollouts using the evolved tools and hooks and skills for assisted tasks. Trajectories from the updated policy guide later proposals. Third, we examine which components should remain at deployment. We evaluate the policy after removing skills and compare performance with and without the evolved tools and hooks.

With a 9B base model, \appname improves pass@1 over GRPO by 9.8, 12.6, and 6.4 percentage points on SWE-bench Verified~\citep{chowdhury2024verified}, Multilingual~\citep{yang2025swesmith}, and Pro~\citep{deng2025swebenchpro}. It also outperforms GRPO training with the human-designed OpenHands harness on all three benchmarks. Under the same base harness, models trained with mechanism-level skills perform best among the three skill-guidance variants when evaluated without skills. Re-evolving tools and hooks as the policy changes outperforms keeping their early versions fixed. Holding the trained policy weights fixed, removing the evolved tools and hooks at evaluation reduces pass@1 by up to 7.7 percentage points. Together, these results show that gains from skills persist after their removal, while the evolved tools and hooks remain necessary for full performance at deployment.

Our contributions are threefold:

\begin{itemize}[leftmargin=*,nosep]

    \item \textbf{Skill design for policy learning.} We compare three skill-guidance variants and find that mechanism-level skills yield the best repair performance after guidance is removed.

    \item \textbf{Tool and hook evolution during reinforcement learning.} We use current-policy trajectories to propose tools and hooks and select changes using both the pass rate and the trainable rate.

    \item \textbf{Training and deploying the model--harness pair.} We show that gains from skill guidance persist after its removal, while retaining the evolved tools and hooks at deployment preserves the complete performance.

\end{itemize}

%% file: sections/preliminaries.tex
\section{Problem Formulation}
\label{sec:prelim}

\paragraph{Agent and harness.}
An agent combines a policy with a harness that provides skill guidance, tools, and runtime feedback. The policy $\pi_\theta$ selects actions from the interaction history, while the harness executes them and returns observations~\citep{yao2023react,yang2024sweagent}. We represent the harness by $(H,s)$, where $s$ is optional skill guidance provided to the policy and $H$ is the runtime configuration. We write
$H=(H_0,\mathcal{T},\mathcal{K})$, where $H_0$ is the base configuration with native Bash, $\mathcal{T}$ is the set of added tools, and $\mathcal{K}$ is the set of runtime hooks. Skills guide repair strategies, tools provide reusable operations, and hooks provide runtime feedback. For task $x$, the policy and harness generate a trajectory
$\tau\sim P(\cdot\mid x,\theta,H,s)$.

\paragraph{Training and deployment.}
Training starts from pretrained weights $\theta_0$ and the base runtime configuration $H_0$. A frozen teacher model generates skills for assisted tasks from current-policy trajectories. A separate teacher model uses these trajectories to propose changes to tools and hooks. Proposed tool and hook changes are evaluated before acceptance. The policy learns from fresh rollouts with this support. At deployment, skills are removed, and the trained policy is paired with the final runtime configuration $H_T$. The deployed agent is therefore represented by $(\theta_T,H_T)$. On unseen tasks, we evaluate
\begin{equation}
J(\theta,H)=
\mathbb{E}_{x\sim\mathcal{D}_{\mathrm{test}}}
\mathbb{E}_{\tau\sim P(\cdot\mid x,\theta,H,\varnothing)}
[R(\tau)].
\label{eq:agent-objective}
\end{equation}
Here, $R(\tau)$ indicates whether the submitted patch passes the task tests. Skill generation and tool and hook selection use only training-side evidence.

%% file: sections/method.tex
\input{figures/overview/figure}

\section{Method: Training the Whole Agent}
\label{sec:method}
\label{sec:method-overview}

Figure~\ref{fig:overview} summarizes \appname. A teacher model uses current-policy trajectories to generate task-specific skills for flawed repair strategies. A separate teacher model proposes changes to tools and hooks for recurring interaction problems across tasks. We accept a proposal only if it improves the observed pass rate or the observed trainable rate without lowering either. The policy then trains through GRPO~\citep{shao2024deepseekmath} on rollouts generated with the adapted harness. Its new trajectories then guide subsequent harness adaptation.

\subsection{Policy-Adaptive Repair Skills}
\label{sec:setup-skills}
\label{sec:skills}

We generate skills from training trajectories to guide the policy's repair strategies. Prior work finds that many skills do not improve software engineering performance~\citep{han2026sweskillsbench}. We therefore compare three ways of generating skills and assess the model's performance after training, with the skills removed.

We first evaluate a skill bank with a fixed capacity, based on prior work that extracts reusable strategies from trajectories~\citep{suzgun2025dynamiccheatsheet,xia2026skillrl}. The teacher reviews trajectories across training tasks and adds, removes, merges, or updates general repair skills. This approach does not consistently improve performance over GRPO, suggesting that these skills are too general to help with individual repair tasks (Section~\ref{sec:skill-abstraction}). A more task-specific alternative is to provide near-answer hints with concrete code changes, including variable, function, or class names. These hints produce shorter trajectories and higher training success, but the trained model performs worse than GRPO when evaluated without them (Section~\ref{sec:skill-abstraction}, Appendix~\ref{app:trajectory-quality}). We therefore propose mechanism-level skills that focus on the policy's flawed repair strategies while leaving the policy to work out the concrete solution.

For each assisted task, the teacher model compares two trajectories from the current policy. Using the issue and the evidence in these trajectories, it identifies a repair strategy and explains when and why it applies. The resulting mechanism-level skill retains the technical reasoning behind the repair but omits repository paths, code symbols, and literal code. This guides the policy's repair strategy while leaving code localization and implementation to the policy. Appendices~\ref{app:skill-examples} and~\ref{app:prompt-outputs} provide generation details and examples.

\subsection{Evolving Tools and Hooks for Policy Training}
\label{sec:setup-evolution}

Training trajectories also reveal recurring problems in how the policy interacts with its environment. The teacher uses this evidence to propose changes to tools and hooks. To support both task performance and policy learning, it uses the pass rate and the trainable rate to guide these proposals. Trainable rollouts are those usable for policy updates (Appendix~\ref{app:candidate-generation}).

The teacher uses current-policy trajectories and the current harness to develop a proposal in four steps.
\emph{(1) Find recurring problems.} The teacher compares trajectories across tasks to identify recurring problems in tool use and feedback.
\emph{(2) Inspect trajectories.} The teacher reads the tool calls, outputs, and hook messages to understand each recurring problem. For example, the policy runs fewer verification tests later in training, even though it has testing tools and receives reminders. The teacher inspects these trajectories to understand why the policy skips tests before submitting a patch.
\emph{(3) Design changes.} To address the identified problems, the teacher can propose one or more tools, hooks, or a combination of both. Tools package repair operations into reusable actions. Hooks give the policy feedback during execution. In the testing example, the teacher modifies the patch-preparation tool so that it automatically runs related tests when preparing a patch. A hook then reports any test failures and reminds the policy to check them before submission.
\emph{(4) Implement the proposal.} The teacher implements the proposed changes as a candidate harness for evaluation.

As the policy learns, its new trajectories guide further adaptation of tools and hooks (Appendix~\ref{app:verification}). Appendix~\ref{app:candidate-generation} provides the full proposal procedure.

\subsection{Training-Aware Harness Selection}
\label{sec:acceptance}

We select harness changes to support both task success and policy learning. GRPO uses reward differences among attempts on the same task to update the policy~\citep{shao2024deepseekmath,yu2025dapo}. Groups in which all attempts succeed or all fail provide no reward contrast for policy updates. Successful rollouts in an all-success group therefore count toward the pass rate without contributing to policy learning. To account for training usability, we also measure the trainable rate, the fraction of sampled rollouts eligible for policy updates (Appendix~\ref{app:candidate-generation}).

We evaluate each proposal by running the candidate and current harnesses with the same policy, tasks, skills, and rollout budget. We accept the proposal only if it improves the observed pass rate or the observed trainable rate without lowering either. These rates are empirical fractions from the finite comparison rollouts; the acceptance rule does not test statistical significance. Accepted changes are used for subsequent policy training. Otherwise, training continues with the current tools and hooks. Changes from a rejected proposal can be proposed again in later rounds.

\paragraph{Policy updates.}
\label{sec:setup-anneal}
After selection, we collect fresh rollouts with the selected tools and hooks, providing skills for assisted tasks. Preparation and comparison rollouts are not reused for training. We update the policy with GRPO using the new rollouts, with loss computed only on policy-generated tokens~\citep{shao2024deepseekmath}.

\paragraph{Evolution across training epochs.}
\label{sec:re-evolution}
Each training epoch begins with up to five rounds of tool and hook evolution. After each round, we update the policy using the selected tools and hooks. These tools and hooks then stay fixed for the remaining training steps in that epoch. The first epoch produces $H_1$. The second epoch starts from $H_1$ and uses trajectories from the updated policy to evolve $H_2$.

%% file: figures/overview/figure.tex
\begin{figure*}[t]
\centering
\includegraphics[width=\textwidth,pagebox=cropbox]{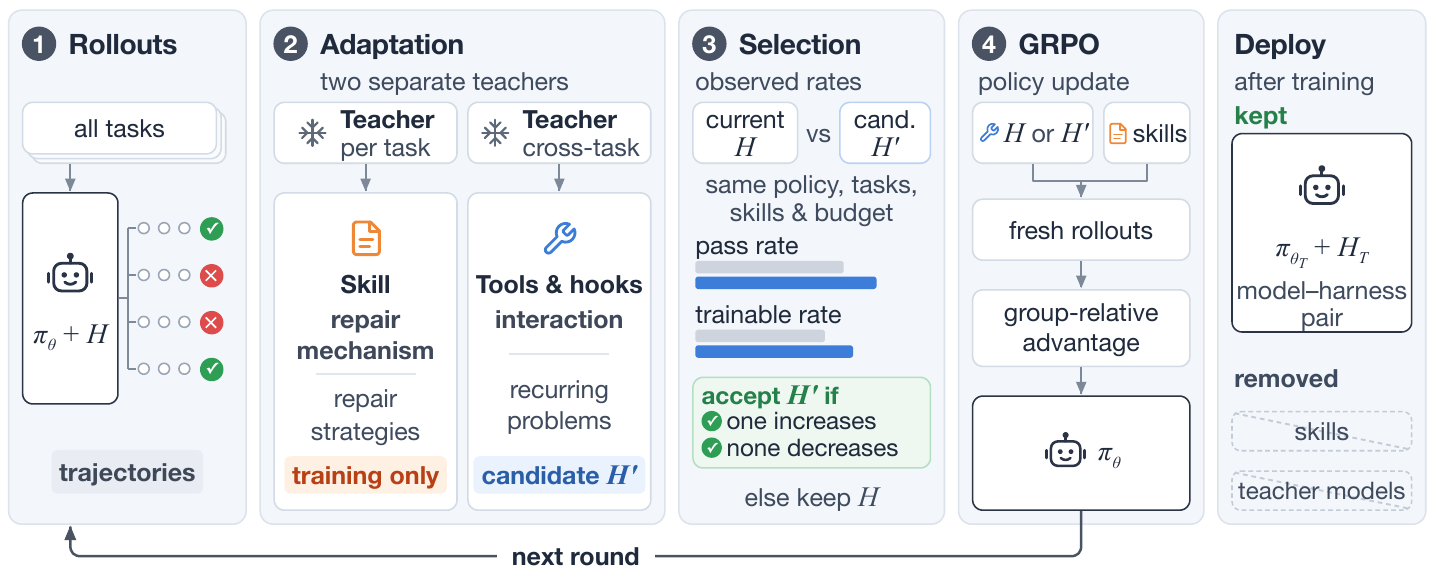}
\caption{Overview of \appname. Two separate teachers derive skills and propose tools and hooks~$H'$ from current-policy trajectories. $H'$ is kept only if it improves the observed pass rate or the observed trainable rate without lowering either. Deployment keeps the trained policy and the final tools and hooks, without skills or teachers.}
\label{fig:overview}
\end{figure*}

%% file: sections/experimental_setup.tex
\section{Experiments}
\label{sec:setup}

\subsection{Experimental Setup}
\label{sec:setup-eval}

\paragraph{Training setup.}
We train on a fixed subset of the R2E-Gym-Lite training split~\citep{jain2025r2e}, shared across all training configurations.
The training tasks are disjoint from the evaluation benchmarks.
In \appname, skills are provided for all training tasks in the first epoch and a fixed subset in the second. Appendix~\ref{app:task-pool} details task selection and the assisted subset.
Our base model is Qwen3.5-9B~\citep{qwen2026qwen35}, and the base harness $H_0$ uses mini-swe-agent~\citep{yang2024sweagent}.
We use the frozen GLM-5.3-Flash teacher~\citep{zai2026glm53flash} to generate skills, and GPT-5.6 Sol~\citep{openai2026gpt56sol} through Codex~\citep{openai2025codex} to propose and implement tools and hooks. Appendix~\ref{app:hyperparameters} records the model identifiers and access interfaces.
Policy training uses GRPO~\citep{shao2024deepseekmath}.
A submitted patch receives a reward of 1 if it passes the task tests and 0 otherwise.
The GRPO baseline receives twice as many policy-training rollouts as the other trained configurations.
Appendix~\ref{app:impl} gives implementation details and hyperparameters, and Appendix~\ref{app:costs} reports the additional costs of preparation, harness comparison, and teacher calls.

\paragraph{Configurations.}
\emph{Base model} uses the initial policy and base harness $H_0$ without training.
\emph{GRPO} trains the policy with $H_0$ and no skills.
\emph{OpenHands-GRPO} uses the human-designed OpenHands harness, also without skills~\citep{wang2025openhands}.

To compare three ways of generating skills, \emph{Skill-Hint}, \emph{Skill-Mechanism}, and \emph{Skill-Bank} use near-answer hints, mechanism-level skills, and general repair skills from a skill bank, respectively.
All three train the policy with the base harness $H_0$.

We compare three further configurations to test whether tools and hooks need to adapt as the policy learns.
All three use mechanism-level skills during policy training.
Our full method, \emph{\appname}, alternates policy updates with tool and hook evolution in both epochs.

\emph{Pre-Evolved} tests whether evolving tools and hooks before policy training is enough.
It follows the same search steps and uses the same tasks as \appname.
However, the policy stays fixed at the base model throughout the search.
It then trains the policy for two epochs with the resulting harness $H_{\mathrm{pre}}$ fixed.

\emph{$H_1$+Mechanism} tests whether tools and hooks need to evolve again in the second epoch.
It shares the first epoch with \appname, so both start the second epoch with the same policy checkpoint and harness $H_1$.
Both continue training the policy, but $H_1$+Mechanism keeps $H_1$ fixed, while \appname evolves it into $H_2$.
Appendix~\ref{app:frozen-h1} reports additional fixed-harness configurations.

\paragraph{Evaluation.}
\label{sec:setup-eval-protocol}
We report pass@1 on SWE-bench Verified~\citep{jimenez2024swebench,chowdhury2024verified}, SWE-bench Multilingual~\citep{yang2025swesmith}, and SWE-bench Pro~\citep{deng2025swebenchpro}.
Each evaluation run uses one attempt per task.
We repeat each evaluation three times and report the mean pass@1.
We assess statistical significance using paired bootstrap confidence intervals for performance differences (Appendix~\ref{app:paired-ci}).
All configurations use the same evaluation limits (Appendix~\ref{app:runtime}).
Evaluation uses no skills, teacher assistance, or further harness updates.
Table~\ref{tab:master} compares complete agents, each evaluated with its deployment harness.

%% file: sections/main_results.tex
\subsection{Main Results}
\label{sec:main-results}

\input{sections/master_table}

\paragraph{\appname performs best on average.}
\appname achieves an average pass@1 of 44.9\%, the highest in Table~\ref{tab:master}, compared with 35.3\% for GRPO.
Its gains over GRPO are 9.8, 12.6, and 6.4 percentage points on Verified, Multilingual, and Pro, respectively.
Both use the same 9B base model, and evaluation uses no skills or teacher calls.

\paragraph{The gains also hold against OpenHands-GRPO.}
OpenHands-GRPO trains the same base model with a human-designed harness instead of $H_0$.
\appname exceeds it by 6.2, 4.0, and 7.4 percentage points on the three benchmarks.
The advantage is therefore not limited to comparison with the minimal base harness.

\paragraph{Mechanism-level skills perform best after removal.}
The three skill variants use the same base harness $H_0$, so they compare different ways of generating skills.
After the skills are removed, the model trained with mechanism-level skills reaches an average score of 38.3\%, compared with 32.7\% for near-answer hints and 34.1\% for skill-bank generation.
Only mechanism-level skills improve performance over GRPO on all three benchmarks.
Near-answer hints achieve higher training success, but this does not lead to better performance without the hints (Section~\ref{sec:skill-abstraction}).

\paragraph{Evolution during training improves performance compared to pre-training search.}
Pre-Evolved uses the same search steps and tasks as \appname, but its policy remains at the base model throughout the search.
Both configurations use mechanism-level skills when training the policy.
\appname scores 2.4, 4.3, and 2.7 percentage points higher on Verified, Multilingual, and Pro.
These results favor using trajectories from the updated policy to improve tools and hooks.

\paragraph{Evolving tools and hooks again in the second epoch improves performance.}
$H_1$+Mechanism and \appname share the first epoch and start the second with the same policy checkpoint and $H_1$.
Both continue training the policy, but only \appname evolves the tools and hooks again.
This raises pass@1 over $H_1$+Mechanism by 2.2, 4.6, and 2.0 percentage points on Verified, Multilingual, and Pro.
Later trajectories show that the policy runs fewer verification tests despite having testing tools and reminders.
The teacher changes the tool that prepares the patch so that it automatically runs related tests. A hook then reports any test failures and reminds the policy to check them before submission (Section~\ref{sec:harness-evolution}).

%% file: sections/master_table.tex
\begin{table*}[t]
\centering
\small
\newcommand{\dgrpo}[1]{{\scriptsize(#1)}}
\setlength{\tabcolsep}{5pt}
\renewcommand{\arraystretch}{1.05}
\begin{tabular*}{\textwidth}{@{\extracolsep{\fill}}l cc ccc c@{}}
\toprule
\textbf{Variant} & \textbf{Skill} & \textbf{Harness} & \textbf{SWE-Verified} & \textbf{SWE-Multilingual} & \textbf{SWE-Pro} & \textbf{Avg.} \\
\midrule
Base model & $\times$ & $H_0$ & 42.2 & 19.3 & 31.8 & 31.1 \\
GRPO & $\times$ & $H_0$ & 47.8 & 24.7 & 33.3 & 35.3 \\
OpenHands-GRPO & $\times$ & OpenHands & 51.4 & 33.3 & 32.3 & 39.0 \dgrpo{+3.7} \\
\midrule
Skill-Hint & Hint & $H_0$ & 45.4 & 22.3 & 30.4 & 32.7 \dgrpo{$-$2.6} \\
Skill-Mechanism & Mechanism & $H_0$ & 52.2 & 28.0 & 34.6 & 38.3 \dgrpo{+3.0} \\
Skill-Bank & Bank & $H_0$ & 48.2 & 20.3 & 33.8 & 34.1 \dgrpo{$-$1.2} \\
\midrule
Pre-Evolved & Mechanism & $H_{\mathrm{pre}}$ & 55.2 & 33.0 & 37.0 & 41.7 \dgrpo{+6.4} \\
$H_1$+Mechanism & Mechanism & $H_1$ & 55.4 & 32.7 & 37.7 & 41.9 \dgrpo{+6.6} \\
\appname & Mechanism & $H_1 \rightarrow H_2$ & \textbf{57.6} & \textbf{37.3} & \textbf{39.7} & \textbf{44.9} \dgrpo{+9.6} \\
\bottomrule
\end{tabular*}
\caption{Pass@1 (\%) on three repair benchmarks. All configurations are evaluated without skills and retain their final harness. Avg.\ is the mean across benchmarks. Parentheses show the difference from GRPO in percentage points.}
\label{tab:master}
\end{table*}

%% file: sections/analysis.tex
\flowpanel[\subsection{Why Mechanism-Level Skills?}
\label{sec:skill-abstraction}
\label{sec:why-skills-help}]{figure}{0.48\textwidth}{%
\centering
\includegraphics[width=\linewidth]{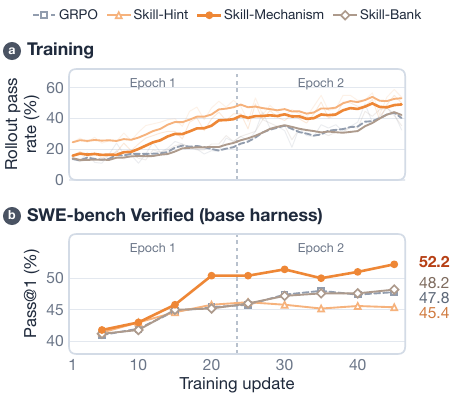}
\caption{Training success and Verified performance without skills. Faint lines show individual updates, bold lines five-update trailing means; dashed lines separate the two training epochs.}
\label{fig:skill-abstraction}
}

Figure~\ref{fig:skill-abstraction} shows that skill-bank training closely follows GRPO in both training success and Verified performance.
This suggests that its general repair strategies offer limited help for individual tasks.
Near-answer hints increase training success substantially, but the trained model performs worse than GRPO after skills are removed.
Mechanism-level skills improve training success while also achieving the best performance without skills.
Higher success during training does not necessarily lead to better performance after training.

To examine this difference, we first compare the rewards of attempts on the same task.
GRPO uses these reward differences to update the policy~\citep{shao2024deepseekmath,yu2025dapo}.
When every attempt succeeds or every attempt fails, the rewards cannot distinguish better attempts.
With mechanism-level skills, 46.5\% of sampled groups contain both successful and failed attempts, compared with 39.0\% with near-answer hints.
The latter produce more groups in which every attempt succeeds (Appendix~\ref{app:reward-groups}).
Their higher success rate therefore also comes with more groups that provide no reward difference for learning.

Providing concrete answers also changes what the policy needs to do during training.
These answers allow it to skip many steps required for independent repair, including locating the relevant code and deciding how to modify it.
Providing concrete answers reduces the average number of tokens per rollout by 40.8\% relative to GRPO, compared with 14.5\% for mechanism-level skills.
The trajectories are much shorter.
However, they can leave out too many steps needed to solve the task independently.
At evaluation, the policy must complete these steps without assistance.
The model trained with concrete answers performs worse than GRPO after skills are removed (Appendix~\ref{app:trajectory-quality}).

\input{sections/harness_analysis}

%% file: sections/harness_analysis.tex
\flowpanel[\subsection{How Do Tools and Hooks Evolve During Training?}
\label{sec:harness-evolution}]{figure}{0.48\textwidth}{%
\centering
\includegraphics[width=\linewidth]{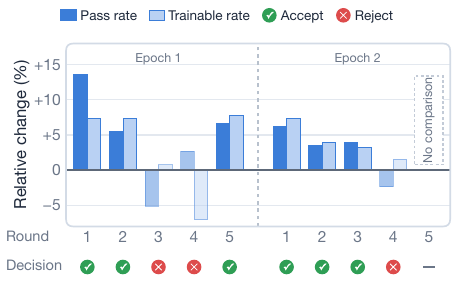}
\caption{Relative changes (\%) in pass rate and trainable rate for each candidate compared with the current harness in the same round.}
\label{fig:harness-evolution}
}

Figure~\ref{fig:harness-evolution} reports the relative changes in pass rate and trainable rate for each candidate compared with the current harness.

\paragraph{Early evolution targets context management and patch submission.}
Reading and search tools limit the text returned in each call, preserving context for later repair steps.
Editing tools apply changes directly, reducing the need for shell commands that can fail because of quoting or syntax errors.
The first proposal containing these tools raises the pass rate by 13.7\% and the trainable rate by 7.4\%.
Later proposals add tools that generate patches and display code changes, helping the policy prepare and check its submission.
Hooks flag repeated actions and remind the policy to reserve steps for checking and submission.
The accepted changes form $H_1$ (Table~\ref{tab:harness-components}).

\paragraph{Later evolution focuses on testing and revising repairs.}
With $H_1$ kept fixed, the policy runs fewer verification tests in the second epoch despite having a testing tool and reminders (Appendix~\ref{app:verification}).
The teacher adds automatic testing to the tool that prepares the patch.
It selects and runs tests related to the changed code.
A hook reports failures and reminds the policy to inspect them before submission.
This proposal raises the pass rate by 6.3\% and the trainable rate by 7.4\%.

Revised budget reminders discourage repeatedly running every test in the repository while still encouraging tests for the current repair.
Another proposal helps the policy undo unsuccessful edits.
When a Git command for undoing changes is blocked, a hook recommends the new \texttt{restore\_file} tool.
This tool restores one file to its original version while preserving edits to other files, so the policy does not have to discard its other work.
Both proposals improve the two rates and become part of $H_2$ (Table~\ref{tab:harness-changes}).

\paragraph{A rejected proposal.}
One proposal adds a hook that asks the policy to compare its patch with the task requirements and identify parts that have not been checked.
It raises the pass rate by 2.7\% but lowers the trainable rate by 7.0\%, so it is rejected.
In a later round, the teacher combines review with \texttt{view\_diff}.
This tool displays the lines added, changed, or deleted in selected files, letting the policy inspect its actual edits during review.
The later proposal improves both rates and is accepted.
A component from a rejected proposal can therefore be reconsidered in a later proposal.
Appendix~\ref{app:harness-evolution} gives the remaining rejected proposals and the complete evolution record.

\flowpanel[\subsection{Why Train and Deploy with the Evolved Harness?}
\label{sec:complete-agent}]{table}{0.52\textwidth}{%
\centering
\small
\newcommand{\gain}[1]{{\scriptsize(+#1)}}
\setlength{\tabcolsep}{1.5pt}
\begin{tabular}{@{}llccc@{}}
\toprule
Model & Harness & Verified & Multi. & Pro \\
\midrule
\multirow{2}{*}{Skill-Mechanism} & $H_0$ & 52.2 & 28.0 & 34.6 \\
 & $H_2$ & 54.2 \gain{2.0} & 33.3 \gain{5.3} & 36.9 \gain{2.3} \\
\midrule
\multirow{2}{*}{\appname} & $H_0$ & 52.8 & 29.6 & 35.0 \\
 & $H_2$ & 57.6 \gain{4.8} & 37.3 \gain{7.7} & 39.7 \gain{4.7} \\
\bottomrule
\end{tabular}
\caption{Pass@1 (\%) without skills under $H_0$ and $H_2$. Parentheses show each model's gain from $H_0$ to $H_2$ in percentage points.}
\label{tab:toolhook-harness-diagnostics}
}

We next examine whether the evolved tools and hooks should remain with the policy at deployment.
Both Skill-Mechanism and \appname use mechanism-level skills during training.
Skill-Mechanism trains with $H_0$, while \appname trains as its tools and hooks evolve.
After training, we keep each policy fixed and evaluate it under both $H_0$ and $H_2$.
All skills are removed for evaluation (Table~\ref{tab:toolhook-harness-diagnostics}).

\paragraph{Adding $H_2$ at evaluation improves performance.}
Skill-Mechanism benefits from $H_2$ on all three benchmarks, although it was trained only with $H_0$.
Its Verified score rises from 52.2\% to 54.2\%.
This improvement comes from changing the tools and hooks, without any additional policy training.

\paragraph{Training with the harness brings larger gains.}
Switching from $H_0$ to $H_2$ at evaluation raises \appname's Verified score by 4.8 percentage points, compared with 2.0 percentage points for Skill-Mechanism.
The same pattern holds on Multilingual and Pro.
The difference between these gains is statistically significant when the three benchmarks are combined (Appendix~\ref{app:paired-ci}).
These results support training the policy with the tools and hooks it will use at deployment.

\paragraph{Removing the evolved tools and hooks lowers performance.}
Replacing $H_2$ with $H_0$ reduces \appname's pass@1 by 4.8, 7.7, and 4.7 percentage points on Verified, Multilingual, and Pro, respectively.
These drops occur without any change to the policy weights.
Retaining the evolved tools and hooks is therefore necessary to preserve the trained agent's full performance.
The deployed model--harness pair consists of the trained policy and these tools and hooks, with skills removed.

%% file: sections/discussion_evolution.tex
\flowpanel[\subsection{Can the Evolved Harness Be Reused for Training?}
\label{sec:frozen-l2}]{figure}{0.48\textwidth}{%
\centering
\includegraphics[width=\linewidth]{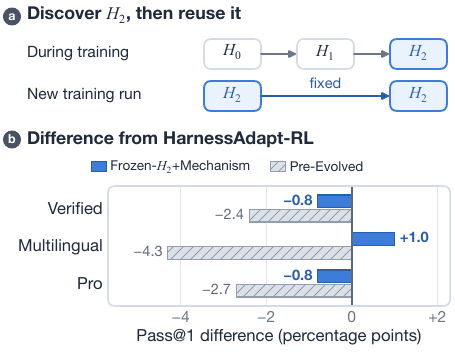}
\caption{Pass@1 differences (percentage points) from \appname. Frozen-$H_2$+Mechanism and Pre-Evolved both train with fixed harnesses and mechanism-level skills, then remove skills at evaluation.}
\label{fig:frozen-harness}
}

We compare reusing $H_2$ with searching for a harness before policy training.
Frozen-$H_2$+Mechanism reuses $H_2$ discovered by \appname.
Pre-Evolved instead searches with the base policy fixed, using the same search steps and tasks as \appname.
After obtaining their harnesses, both variants train a new policy from the same base model.
Their tools and hooks remain fixed throughout training.
We provide mechanism-level skills on the same schedule as in the full method (Section~\ref{sec:setup-eval}).

Frozen-$H_2$+Mechanism reaches 56.8\%, 38.3\%, and 38.9\% on Verified, Multilingual, and Pro, respectively.
These scores are within 1.0 percentage point of \appname on every benchmark (Figure~\ref{fig:frozen-harness}).
They also exceed Pre-Evolved by 1.6, 5.3, and 1.9 percentage points.
Thus, reusing the harness discovered during policy training gives better results than using the harness found before training.

Frozen-$H_2$+Mechanism achieves similar scores to \appname.
However, it reuses a harness that \appname has already discovered during policy training.
In our experiments, searching with the fixed base policy does not produce a harness with the same performance.
The strong reuse result therefore depends on first discovering $H_2$ through evolution during training.
Once $H_2$ is available, this evolution does not need to be repeated in the new training run.

%% file: sections/related_work.tex
\section{Related Work}
\label{sec:related}

\paragraph{Skills and policy learning.}
Voyager builds reusable skills from interaction~\citep{wang2024voyager}, while SkillRL combines skill evolution with reinforcement learning~\citep{xia2026skillrl}.
Agentic Context Engineering (ACE) maintains a playbook of strategies and updates it using feedback from the agent's interactions~\citep{zhang2026agentic}.
Socratic-SWE uses skills derived from trajectories to generate training tasks~\citep{xiao2026socraticswe}.
Hindsight Hint Distillation trains on trajectories generated with hints, while PATS adjusts guidance to the current policy~\citep{hhd2026,pats2026}.
Both remove this guidance at deployment.
We compare near-answer hints, mechanism-level skills, and general skills from a skill bank in software repair.
Our comparison focuses on the trained model's performance after skills are removed.

\paragraph{Harness design and optimization.}
Tool interfaces and feedback affect agent behavior~\citep{yang2024sweagent,shinn2023reflexion,madaan2023selfrefine}.
Meta-Harness automates harness search using earlier candidates and their execution trajectories~\citep{lee2026metaharness}.
Retrospective Harness Optimization (RHO) updates skills and tools from past trajectories, selecting changes through the agent's own preferences without external grading~\citep{pan2026evolving}.
Other methods use revision history, reference trajectories, or interactions across tasks to guide changes~\citep{lee2026rhi,jiang2026harnessevolve,du2026livingharness}.
These methods keep model weights fixed during harness optimization.
We evolve tools and hooks during policy training.
The teacher uses trajectories from the updated policy to propose changes for newly observed problems.
Our experiments compare this approach with search under a fixed base policy.

\paragraph{Policy training and harness adaptation.}
CoEvolve adapts training data alongside the policy~\citep{yang2026coevolve}, while SEARL jointly optimizes the policy and tool graph memory~\citep{feng2026searl}.
ClawGym~II studies reinforcement learning with complex harnesses~\citep{clawgym2026}.
Co-Harness alternates harness updates with supervised fine-tuning (SFT) on trajectories generated by the improved harness~\citep{chen2026coharness}.
\appname combines GRPO with tool and hook selection based on the pass rate and trainable rate.
A proposal is accepted only if at least one observed rate improves and neither decreases.
We also examine which components are needed at deployment.
We test whether skill gains persist after removal and how removing the evolved tools and hooks affects performance.

%% file: sections/conclusion.tex
\section{Conclusion}
\label{sec:conclusion}

We introduced \appname, which trains coding agents with GRPO and adapts their harnesses from training trajectories.
It generates task-specific skills and evolves tools and hooks as the policy learns.
A proposal is accepted only if it improves the observed pass rate or the observed trainable rate without lowering either.
With a 9B base model, \appname improves pass@1 over GRPO by 9.8, 12.6, and 6.4 percentage points on SWE-bench Verified, Multilingual, and Pro.
It also outperforms OpenHands-GRPO.
Mechanism-level skills outperform near-answer hints and general skills from a skill bank after skills are removed.
Evolving tools and hooks during policy training performs better than searching only before training or keeping the first-epoch harness fixed.
Once discovered, the evolved harness can support a new training run without further evolution.
Removing its tools and hooks at deployment lowers pass@1 by up to 7.7 percentage points.
Training therefore produces a model--harness pair: the trained policy and the evolved tools and hooks, with task-specific skills removed at deployment.

%% file: sections/data_availability.tex
\section*{Data Availability}

The model weights, evolved harness, training code, and data are available at \url{https://github.com/lingxiaotang/HarnessAdapt-RL}.

%% file: appendix/limitations.tex
\section*{Limitations}
\label{app:limits}

All tasks are software repair from SWE-bench-style benchmarks, and the skills, tools, and hooks were derived for this setting. We have not tested whether the same procedure produces useful support in other agent domains, such as web navigation or data analysis. We also use a single 9B base model. We do not know whether larger models, which may already handle many of the operations that our tools package, would benefit from harness evolution to the same extent, or whether the preferred form of guidance would stay the same across model families.

Each configuration is trained once. The three repetitions are evaluations of a fixed trained checkpoint, not independent training runs (Appendix~\ref{app:protocol-status}). The reported confidence intervals characterize task-level variation for these checkpoints; they do not quantify variability across training seeds or independent harness-search runs.

Candidate selection uses observed pass and trainable rates from a finite rollout comparison (Appendix~\ref{app:candidate-generation}). Its acceptance rule is not a significance test: an observed improvement need not imply an improvement in the underlying success or trainability probabilities. Rollout noise and repeated candidate search can therefore affect which changes are retained. The statistical analysis of the final evaluations does not remove this uncertainty from the search process.

%% file: appendix/data_harness.tex
\section{Data and Harness Specifications}
\label{app:data}

\sidepanel[\subsection{Task Pool and Assisted Subset}
\label{app:task-pool}]{table}{0.52\textwidth}{%
\centering
\small
\renewcommand{\arraystretch}{1.08}
\setlength{\tabcolsep}{3.5pt}
\begin{tabular}{@{}lrr@{\hspace{1.2em}}lrr@{}}
\toprule
\textbf{Repository} & \textbf{Tasks} & \textbf{Share} & \textbf{Repository} & \textbf{Tasks} & \textbf{Share} \\
\midrule
pandas & 399 & 37.71\% & tornado & 62 & 5.86\% \\
numpy & 163 & 15.41\% & scrapy & 51 & 4.82\% \\
orange3 & 124 & 11.72\% & datalad & 31 & 2.93\% \\
pillow & 122 & 11.53\% & coveragepy & 24 & 2.27\% \\
aiohttp & 76 & 7.18\% & pyramid & 6 & 0.57\% \\
\midrule
\multicolumn{3}{@{}l}{\textbf{Total}} & & \textbf{1{,}058} & \textbf{100\%} \\
\bottomrule
\end{tabular}
\caption{Repository distribution of the training pool.}
\label{tab:task-pool}
}

We draw the training pool from the train split of R2E-Gym-Lite~\citep{jain2025r2e}, which contains 4{,}578 tasks.
We first remove tasks with a pass@4 of 100\%, then randomly sample 1{,}058 of the remaining tasks and use this pool for every training configuration.
Table~\ref{tab:task-pool} reports its repository distribution.

The 512-task assisted subset is drawn uniformly at random from the pool, without reference to task difficulty, and is held fixed across configurations.
For configurations with skills, the first epoch assists all 1{,}058 tasks and the second assists only this subset. Skill-Gated further restricts guidance to tasks whose preparation attempts all fail.
Tasks outside the subset keep the current harness but receive no skill.
We deduplicate the training pool against SWE-bench Verified, Multilingual, and Pro and find no overlap.
Tasks are visited in the same order in every configuration.

\sidepanel[\subsection{Skill Generation}
\label{app:skill-examples}]{table}{0.52\textwidth}{%
\centering
\small
\setlength{\tabcolsep}{4pt}
\renewcommand{\arraystretch}{1.15}
\begin{tabular}{@{}L{.22\linewidth}L{.37\linewidth}L{.33\linewidth}@{}}
\toprule
\textbf{Input} & \textbf{Visible to the teacher} & \textbf{Use} \\
\midrule
Issue & Yes, as the user message in both trajectories & States what the repair must accomplish \\
Trajectories & Two full preparation trajectory records, with outcomes & Behavioral comparison \\
Tests & Only commands and outputs inside the two trajectories & Evidence the policy actually observed \\
\midrule
Direct repository access & No & -- \\
Hidden tests, grader artifacts & No direct access & -- \\
\bottomrule
\end{tabular}
\caption{Teacher inputs for skill generation.}
\label{tab:teacher-inputs}
}

\paragraph{Teacher inputs.}
Each assisted task visit starts a skill-generation request, which returns a Markdown response.
The teacher sees the issue and trajectory records, including any source code and test output shown in those records.
It has no direct repository access and cannot inspect hidden tests or grader artifacts (Table~\ref{tab:teacher-inputs}).
Only the resulting skill is shown to the policy, and neither the teacher nor any training skill is used at evaluation.
For mechanism-level skills, the skill describes the repair mechanism rather than reproducing the diff.

\paragraph{Trajectory pairing.}
The teacher compares two of the four preparation rollouts of a task, preferably one success and one failure.
When all rollouts succeed, the pair consists of two successes and the skill describes the shared successful mechanism.
When all rollouts fail, the pair consists of two failures, and the teacher decides from the issue and the partial progress in the trajectories which repair direction the skill describes.
The skill format is the same in all three cases.

\paragraph{Guidance variants.}
The three variants differ in both specificity and evidence scope.
Near-answer hints name the file, function, and variables involved and state the concrete edit.
Each mechanism-level skill describes a single repair mechanism, with repository names, paths, symbols, literal code, and diff text removed.
Both are generated from a single task, using its issue and the two trajectories with the test outputs they contain.
Skill-bank guidance instead summarizes general procedures that recur across tasks.
The teacher reads the task reports from the current batch together with the previous meta-skill bank and rewrites the bank into at most ten procedures within 1{,}000 English words.
Evidence identifiers are removed before the bank is shown to the policy.
Section~\ref{sec:skill-abstraction} compares the training and evaluation results of these variants.
For Skill-Mechanism, only the Task-specific Skill section of the teacher's response is injected.
Appendix~\ref{app:prompt-skill} gives the full prompts for all three variants.

\paragraph{Representative outputs.}
Appendix~\ref{app:prompt-outputs} shows representative teacher outputs for each guidance type.

\paragraph{Length and leakage control.}
Excluding the three field labels, mechanism-level skills average 100.67 whitespace-delimited words, with a median of 100 and a maximum of 154.
We screen mechanism-level skills for answer-locating paths, symbols, and code.
Near-answer hints intentionally include such details and are not subject to this restriction.
A mechanism-level skill that fails the screen is regenerated up to three times.
If it still fails, the task is trained without a skill.

\paragraph{Second epoch and Skill-Gated.}
Skills for the second epoch are regenerated rather than reused from the first.
Skill-Gated invokes the teacher only when all four preparation attempts of a task fail, and generates the skill from two of these failed trajectories chosen at random.

\subsection{Execution Limits and Submission}
\label{app:runtime}

All configurations share the same execution budgets.
Training rollouts allow a 64K context (65{,}536 tokens) and at most 60 turns, while evaluation allows a 128K context (131{,}072 tokens) and at most 150 turns. Both use temperature 1.0.
The shorter training budget bounds the cost of collecting multiple rollouts per task for each GRPO update.
Evaluation makes a single attempt per held-out task, so the longer budget leaves the trained policy room to read, verify, and submit.
Prior SWE agents also use shorter context or turn budgets during training than at evaluation~\citep{yang2025kimidev,cao2025skyrlagent}.
Shorter interaction horizons can also improve reinforcement-learning stability~\citep{horizon2026}.

\paragraph{Truncation and submission.}
A trajectory is truncated if it exceeds either the turn limit or the context limit, and is counted once if it exceeds both.
A submission made on the final allowed turn counts as complete.
The \texttt{prepare\_patch} tool writes the selected source changes to \texttt{patch.txt} but neither submits nor scores the repair.
Submission is a separate Bash action:
\begin{lstlisting}[style=promptcode]
echo COMPLETE_TASK_AND_SUBMIT_FINAL_OUTPUT && cat patch.txt
\end{lstlisting}
The submitted patch is then applied and tested in a fresh task container.

%% file: appendix/implementation.tex
\section{Implementation and Reproducibility}
\label{app:impl}

\subsection{Hyperparameters and Schedule}
\label{app:hyperparameters}

Policy training uses GRPO implemented in slime~\citep{slime2026}, with a constant learning rate of $10^{-6}$.
GLM-5.3-Flash~\citep{zai2026glm53flash} generates skills through OpenCode~\citep{opencode2026}, and GPT-5.6 Sol~\citep{openai2026gpt56sol} designs and implements tool and hook proposals through Codex~\citep{openai2025codex}.
The release configuration records the provider model identifiers \texttt{glm-5.3-flash} and \texttt{gpt-5.6-sol}.
The setup guide describes the provider endpoints and credentials used by the two CLIs; inference with the released model--$H_2$ pair requires neither teacher.
These identifiers do not pin immutable teacher snapshots, and the historical CLI package versions were not preserved. The reproduction setup therefore requires explicit CLI versions to be supplied and records the installed versions.

Training runs for two epochs over the 1{,}058-task pool.
Each epoch consists of 23 updates of 46 tasks ($23\times 46=1{,}058$), so every batch is full and no task is repeated or padded within an epoch.
Updates 1--23 form the first epoch and updates 24--46 the second, for 2{,}116 task visits in total.
The first epoch assists all tasks, and the second assists the fixed 512-task subset with regenerated skills.

GRPO samples 8 rollouts per task, and every other trained configuration samples 4.
Each update therefore uses 368 training rollouts for GRPO and 184 for the other configurations, or 16{,}928 and 8{,}464 rollouts over the full schedule.
Table~\ref{tab:hyperparameters} lists the remaining settings.

\begin{compacttable}[h]
\centering
\small
\setlength{\tabcolsep}{4pt}
\renewcommand{\arraystretch}{1.1}
\begin{tabular}{@{}L{.38\columnwidth}L{.58\columnwidth}@{}}
\toprule
\textbf{Setting} & \textbf{Value} \\
\midrule
\tabgroup{Schedule}
Policy & Qwen3.5-9B \\
Batch & 46 task groups per update \\
Updates & 23 per epoch, 46 in total \\
Task visits & 1{,}058 per epoch, 2{,}116 in total \\
Rollouts per task & 8 for GRPO, 4 otherwise \\
Rollouts per update & 368 for GRPO, 184 otherwise \\
Training rollouts & 16{,}928 for GRPO, 8{,}464 otherwise \\
\midrule
\tabgroup{Optimization}
Optimizer & Adam \\
Adam $(\beta_1,\beta_2,\epsilon)$ & $(0.9,\ 0.98,\ 10^{-8})$ \\
Weight decay & 0.1 \\
Learning rate & $1\times 10^{-6}$, constant \\
Advantage & Group-mean centering, no std.\ normalization \\
PPO clip & 0.2 (lower), 0.28 (upper) \\
KL / entropy coef. & 0 / 0 \\
Gradient clip & 1.0 \\
\midrule
\tabgroup{Rollout}
Context & 64K tokens (65{,}536) \\
Turns & At most 60 \\
Temperature & 1.0 \\
\midrule
\tabgroup{System}
Precision & BF16 \\
Parallelism & TP 2, CP 2, PP 1 \\
Hardware & 8 GPUs \\
\bottomrule
\end{tabular}
\caption{Training hyperparameters.}
\label{tab:hyperparameters}
\end{compacttable}

All trained configurations share the task pool, task order, batch size, number of updates, optimizer, learning rate, clip and KL settings, loss mask, and the training and evaluation budgets.
They differ only in rollout width, guidance, and harness.
GRPO uses the base harness without skills.
Skill-Hint uses near-answer hints, Skill-Mechanism uses mechanism-level skills, and Skill-Bank uses general procedures from a skill bank. All three keep the base harness.
Skill-Gated adds mechanism-level skills only when all four preparation attempts fail.
OpenHands-GRPO trains without skills under the OpenHands harness and keeps it at evaluation.
Pre-Evolved first keeps the base policy fixed and evolves tools and hooks on the task batches corresponding to the first five updates of each epoch of \appname.
After this search, it trains the policy with mechanism-level skills while keeping $H_{\mathrm{pre}}$ fixed.
Frozen-$H_1$ trains without skills under $H_1$, fixed from the first update.
\appname combines mechanism-level skills with tool and hook evolution in both epochs.
The loss is computed only on policy-generated tokens. Skill text, hook messages, and tool outputs are masked out.
Preparation rollouts inform skill generation and harness proposals and are excluded from optimization.

\subsection{Candidate Generation}
\label{app:candidate-generation}

In each evolution round, the teacher receives all preparation trajectories from that round, the current harness, and the outcomes of earlier proposals, without subsampling.
Within a single call, it analyzes the trajectory evidence, selects a supported problem, and designs and implements one candidate to address it.
It returns the diagnosis, design rationale, and concrete code or edits, or \texttt{no\_change} when the evidence or available interfaces do not support a useful modification.
A bundle may contain several tools and hooks, with corresponding usage instructions, when they serve the same repair workflow.
Appendix~\ref{app:prompt-harness} gives the full proposal prompt.

The generated code is used without human modification.
Before any rollout comparison, we check the candidate's output format, tool schemas, module loading, and a software smoke test.
Tools and hooks outside the proposal must remain byte-identical to the current harness, and task-specific skill text must be unchanged.
A candidate that fails any check is discarded without comparison.

The comparison follows Section~\ref{sec:acceptance}.
Both harnesses are run on fresh rollouts under the same policy, tasks, skills, and rollout budget.
The trainable rate is the share of sampled rollouts eligible for policy updates, and the pass rate is the share whose patch passes the task tests.
We exclude truncated rollouts from the trainable-rate count, motivated by the loss masking of overlong trajectories in prior LLM and agent RL work~\citep{yu2025dapo,luo2025deepswe,cao2025skyrlagent}.
For this metric, we additionally require the remaining rollouts in each task group to include both successful and failed attempts.
When this condition holds, the remaining rollouts are eligible, subject to the unknown-reward exclusion in Appendix~\ref{app:rollout-accounting}.
If the remaining attempts all succeed or all fail, none are trainable.
The denominator for the trainable rate includes all sampled rollouts, including those excluded by these rules.
Successful rollouts in an all-success group count toward the pass rate but not the trainable rate.
Rollouts with unknown rewards count as non-passing and remain in the denominator.
A candidate is accepted if neither observed rate decreases and at least one strictly increases. If both remain unchanged, it is rejected.

Each accepted bundle updates the retained harness.
The first-epoch harness $H_1$ contains seven tools and four hooks (Table~\ref{tab:harness-components}).
The second-epoch harness $H_2$, used by \appname at update 45, contains eight tools and five hooks (Table~\ref{tab:harness-changes}).
Each training epoch begins with at most five harness-evolution rounds. A rejected candidate or a \texttt{no\_change} response also uses one round. The retained harness is fixed for the remaining policy training in that epoch.

\sidepanel[\subsection{Rollout Accounting}
\label{app:rollout-accounting}]{table}{0.52\textwidth}{%
\centering
\small
\setlength{\tabcolsep}{3pt}
\renewcommand{\arraystretch}{1.12}
\begin{tabular}{@{}llccc@{}}
\toprule
\textbf{Configuration} & \textbf{Updates} & \textbf{Prep.} & \textbf{Comp.} & \textbf{Train} \\
\midrule
\multirow{2}{*}{Skill variants} & 1--23 & 184 & 0 & 184 \\
 & 24--46 & $4n_{\mathrm{a}}$ & 0 & 184 \\
\midrule
\multirow{2}{*}{\appname} & 1--23 & 184 & 0 or 368 & 184 \\
 & 24--46 & $4n_{\mathrm{a}}$ & 0 or 368 & 184 \\
\midrule
GRPO & 1--46 & 0 & 0 & 368 \\
Frozen-$H_1$ & 1--46 & 0 & 0 & 184 \\
\bottomrule
\end{tabular}
\caption{Sampled rollouts per update. $n_{\mathrm{a}}$ is the number of assisted tasks in the batch. For \appname, a candidate comparison adds 368 rollouts, otherwise the comparison count is zero.}
\label{tab:rollout-accounting}
}

Table~\ref{tab:rollout-accounting} reports rollouts per update.
Training rollouts follow Appendix~\ref{app:hyperparameters}.
Preparation rollouts are collected only for assisted tasks, four per task, giving 4{,}232 in the first epoch and 2{,}048 in the second.
\appname additionally collects comparison rollouts in rounds that propose a candidate.
Each comparison uses 184 fresh rollouts per harness on the same 46 tasks, with four rollouts per task. Comparison rollouts are discarded once the decision is made.
Rounds that return \texttt{no\_change} require no comparison.

Failed requests are not replaced: an update keeps its scheduled rollouts and does not resample to fill a group.
A rollout with an unknown reward is neither a pass nor a confirmed failure.
We exclude its group from the policy update instead of assigning a reward of 0.
Teacher retries that repair a skill's format do not add training rollouts.
In the skill-training run, training rollouts record 14 summary errors, 11 unknown rewards, and 3 timeouts, and first-epoch preparation rollouts record 6 summary errors and 6 unknown rewards.
Because these events overlap, we do not sum them into a single failure count.

\subsection{Training Runs and Evaluation}
\label{app:protocol-status}
\label{app:checkpoints}

Each configuration is trained once.
We evaluate the checkpoint from update 45, the last saved checkpoint before training ends at update 46.
Checkpoints are saved every five updates, and this choice was made before evaluation.
Each evaluation is repeated three times, with one attempt per task in each run.
We report the mean pass@1, keeping failed or unfinished attempts in the denominator as incorrect.
Appendix~\ref{app:paired-ci} describes the confidence intervals.

Learning curves evaluate Verified every five updates under the same protocol.
Their final points equal the reported scores in Tables~\ref{tab:master} and~\ref{tab:reference-configurations}. Avg.\ is the unweighted mean of the three reported benchmark scores.
The training statistics in Appendices~\ref{app:reward-groups} and~\ref{app:trajectory-quality} cover all 46 updates of the same runs.
They describe sampled groups and rollouts, not unique tasks or independent training replicates.

%% file: appendix/additional_results.tex
\section{Additional Results}
\label{app:extra}

\sidepanel[\subsection{Reward Groups}
\label{app:reward-groups}]{figure}{0.52\textwidth}{%
\begin{minipage}{\linewidth}
\captionsetup{type=table}
\centering
\small
\setlength{\tabcolsep}{4pt}
\renewcommand{\arraystretch}{1.1}
\begin{tabular}{@{}lrrr@{}}
\toprule
\textbf{Configuration} & \textbf{All fail} & \textbf{Mixed} & \textbf{All pass} \\
\midrule
GRPO & 1{,}076 \scriptsize(50.9) & 876 \scriptsize(41.4) & 164 \scriptsize(7.8) \\
Skill-Hint & 710 \scriptsize(33.6) & 825 \scriptsize(39.0) & 581 \scriptsize(27.5) \\
Skill-Mechanism & 841 \scriptsize(39.7) & \textbf{984} \scriptsize(\textbf{46.5}) & 291 \scriptsize(13.8) \\
Skill-Bank & 1{,}064 \scriptsize(50.3) & 876 \scriptsize(41.4) & 176 \scriptsize(8.3) \\
Skill-Gated & 1{,}071 \scriptsize(50.6) & 873 \scriptsize(41.3) & 172 \scriptsize(8.1) \\
\bottomrule
\end{tabular}
\caption{Sampled reward-group counts before training filters, with shares (\%) in parentheses. Each configuration has 2{,}116 groups over updates 1--46.}
\label{tab:reward-counts}
\end{minipage}
\par\vspace{8pt}
\begin{minipage}{\linewidth}
\captionsetup{type=figure}
\centering
\includegraphics[width=0.85\linewidth]{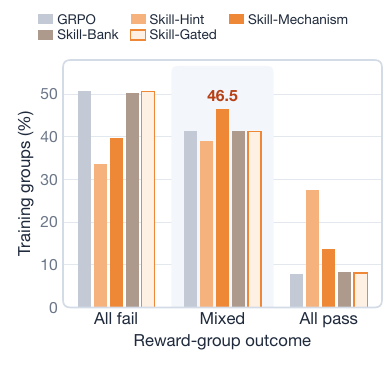}
\caption{Share of sampled groups by reward outcome before training filters. Skill-Mechanism has the largest share of mixed groups.}
\label{fig:reward-groups}
\end{minipage}
}

Section~\ref{sec:why-skills-help} discusses the reward composition of sampled rollout groups.
A group contains the rollouts for one task visit.
Table~\ref{tab:reward-counts} and Figure~\ref{fig:reward-groups} classify the sampled outcomes before training filters are applied.
Each configuration has 2{,}116 groups over updates 1--46.
GRPO samples eight rollouts per group, while the skill variants sample four.
Group size affects the chance of observing mixed outcomes, so comparisons among the skill variants have the same group size, whereas comparisons with GRPO do not.
For this descriptive tally, unknown rewards are placed in the failure category: there are 12, 10, 4, 3, and 6 for GRPO, Skill-Hint, Skill-Mechanism, Skill-Bank, and Skill-Gated, respectively.
Their groups are excluded from policy updates (Appendix~\ref{app:rollout-accounting}).
Thus, these sampled-group counts differ from the groups retained for learning.

Skill-Hint has the largest all-pass share among the skill variants (27.5\%), while Skill-Mechanism has the largest mixed-outcome share (46.5\%).
Skill-Bank and Skill-Gated have distributions close to GRPO's, although their groups contain fewer rollouts.
After training filters are applied, mixed outcomes provide the reward contrast used by GRPO~\citep{shao2024deepseekmath,yu2025dapo}.
The sampled mixed-group share alone does not measure how much useful training signal remains.

\sidepanel[\subsection{Trajectory Length}
\label{app:trajectory-quality}]{figure}{0.52\textwidth}{%
\begin{minipage}{\linewidth}
\captionsetup{type=table}
\centering
\small
\setlength{\tabcolsep}{4pt}
\renewcommand{\arraystretch}{1.1}
\begin{tabular}{@{}lrrr@{}}
\toprule
\textbf{Metric} & \textbf{GRPO} & \textbf{Skill-Hint} & \textbf{Skill-Mechanism} \\
\midrule
\tabgroup{Tokens}
All & 32{,}212 & 19{,}082 & 27{,}535 \\
Successful & 24{,}592 & 14{,}864 & 20{,}828 \\
Failed & 34{,}988 & 22{,}139 & 31{,}316 \\
\midrule
Turns & 52.19 & 34.16 & 47.83 \\
\bottomrule
\end{tabular}
\caption{Mean training-trajectory length over updates 1--46.}
\label{tab:trajectory-length}
\end{minipage}
\par\vspace{8pt}
\begin{minipage}{\linewidth}
\captionsetup{type=figure}
\centering
\includegraphics[width=0.85\linewidth]{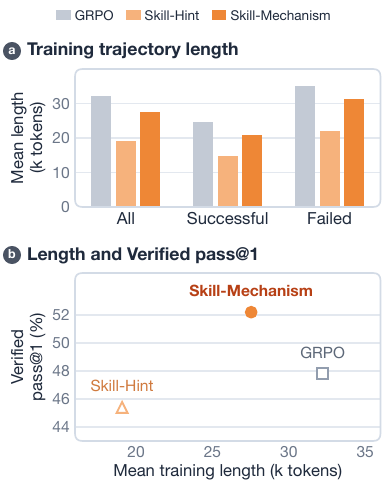}
\caption{(a) Mean training-trajectory length. (b) Mean length against Verified pass@1 without skills under the base harness.}
\label{fig:trajectory-compactness}
\end{minipage}
}

Table~\ref{tab:trajectory-length} reports the mean length of training trajectories over updates 1--46, measured in tokens and interaction turns.
Successful and failed subsets are defined within each configuration and are not matched by task.
Figure~\ref{fig:trajectory-compactness} relates mean length to Verified performance without skills.

Relative to GRPO, Skill-Mechanism shortens trajectories by 14.5\% overall, by 15.3\% for successful rollouts, and by 10.5\% for failed ones.
Skill-Hint shortens them by 40.8\% overall and by 39.6\% and 36.7\% in the two subsets, yet performs worse than GRPO without skills.
Shorter training trajectories are therefore not evidence of better learning on their own.
Because aggregate means mix changes in redundancy, verification, and outcome composition, we do not attribute the reduction to a specific behavior.

\sidepanel[\subsection{Verification Over Training}
\label{app:verification}]{table}{0.52\textwidth}{%
\centering
\small
\setlength{\tabcolsep}{3pt}
\renewcommand{\arraystretch}{1.1}
\begin{tabular}{@{}llrrr@{}}
\toprule
\textbf{Configuration} & \textbf{Harness} & \textbf{Epoch 1} & \textbf{Epoch 2} & \shortstack{\textbf{Relative}\\\textbf{change}} \\
\midrule
GRPO & $H_0$ & 2.87 & 1.58 & $-45\%$ \\
Skill-Mechanism & $H_0$ & 3.16 & 1.78 & $-44\%$ \\
$H_1$+Mechanism & $H_1$ & 2.83 & 1.21 & $-57\%$ \\
\bottomrule
\end{tabular}
\caption{Mean test executions per training trajectory over the first ten updates of each epoch. The change is relative to the first-epoch mean. Harness denotes the harness used in the second epoch.}
\label{tab:verification}
}

Table~\ref{tab:verification} reports the mean number of test executions per trajectory over the first ten updates of each epoch.
For GRPO and Skill-Mechanism, both trained under $H_0$, this count falls by 45\% and 44\%, respectively.
Verification therefore also declines without an evolved harness.
For $H_1$+Mechanism, which retains $H_1$ in the second epoch, the count falls by 57\%.
The existing checking tools and reminders do not prevent this observed decline, motivating further harness changes.
These are descriptive statistics from one training run per configuration and do not establish the cause of the decline.

\subsection{Paired Confidence Intervals}
\label{app:paired-ci}

Each configuration is trained once, and each checkpoint is evaluated three times.
Following the task-level comparison in \citet{miller2024errorbars}, we first average the three outcomes for each task and configuration.
We then resample tasks with replacement, using the same sampled tasks for both configurations, and recompute their performance difference.
Table~\ref{tab:paired-ci} reports the resulting 95\% confidence intervals.
Pooled results cover all 1{,}442 tasks and weight each benchmark by its number of tasks.
To compare the two policies' gains from $H_2$, we first subtract each policy's $H_0$ score from its $H_2$ score, then compare those gains within each resample.
The intervals describe variation across tasks for these trained checkpoints and averaged evaluation outcomes.
They do not estimate variation from retraining the models.

\paragraph{Re-evolution and reuse.}
\appname exceeds $H_1$+Mechanism by 2.6 percentage points pooled, with a confidence interval of [+0.9, +4.3].
The individual intervals exclude zero on Multilingual but include zero on Verified and Pro.
The pooled gains over GRPO, Skill-Mechanism, and OpenHands-GRPO also exclude zero.
Frozen-$H_2$+Mechanism exceeds $H_1$+Mechanism by 2.2 percentage points pooled ([+0.4, +3.9]), providing additional evidence that the discovered $H_2$ remains useful in a new training run.
Frozen-$H_1$+Mechanism and $H_1$+Mechanism have a pooled difference of 0.6 percentage points ([$-$1.1, +2.3]).
These reuse comparisons change the harness schedule and are not repeated seeds of the same configuration.

\begin{table*}[t]
\centering
\small
\newcommand{\cigroup}[1]{\multicolumn{5}{@{}l}{\textit{#1}}\\}
\setlength{\tabcolsep}{1.5pt}
\renewcommand{\arraystretch}{1.1}
\begin{tabular*}{\textwidth}{@{\extracolsep{\fill}}L{.32\textwidth}cccc@{}}
\toprule
\textbf{Comparison} & \textbf{Verified} & \textbf{Multilingual} & \textbf{Pro} & \textbf{Pooled} \\
\midrule
\cigroup{Main results}
\appname $-$ GRPO & +9.8 [+6.3, +13.3] & +12.6 [+8.7, +16.6] & +6.4 [+3.7, +9.1] & +8.9 [+7.0, +10.7] \\
\appname $-$ $H_1$+Mechanism & +2.2 [$-$0.9, +5.3] & +4.6 [+1.3, +7.9] & +2.0 [$-$0.5, +4.5] & +2.6 [+0.9, +4.3] \\
$H_1$+Mechanism $-$ Frozen-$H_1$ & +2.2 [$-$0.8, +5.3] & +1.4 [$-$2.2, +4.8] & +2.0 [$-$0.5, +4.6] & +1.9 [+0.2, +3.7] \\
Frozen-$H_1$ $-$ GRPO & +5.4 [+1.9, +8.8] & +6.6 [+2.4, +10.9] & +2.4 [$-$0.3, +5.2] & +4.3 [+2.3, +6.3] \\
\appname $-$ Skill-Mechanism & +5.4 [+2.0, +8.9] & +9.3 [+5.4, +13.2] & +5.1 [+2.2, +8.0] & +6.1 [+4.1, +8.0] \\
\appname $-$ OpenHands-GRPO & +6.2 [+2.6, +9.7] & +4.0 [$-$0.3, +8.3] & +7.4 [+4.6, +10.2] & +6.3 [+4.3, +8.2] \\
OpenHands-GRPO $-$ GRPO & +3.6 [+0.3, +6.9] & +8.6 [+4.6, +12.7] & $-$1.0 [$-$3.9, +1.9] & +2.6 [+0.7, +4.6] \\
Frozen-$H_1$ $-$ OpenHands-GRPO & +1.8 [$-$1.5, +5.1] & $-$2.0 [$-$6.3, +2.3] & +3.4 [+0.5, +6.2] & +1.7 [$-$0.2, +3.7] \\
\midrule
\cigroup{Skill variants and coverage}
Skill-Mechanism $-$ GRPO & +4.4 [+1.1, +7.7] & +3.3 [$-$0.1, +7.0] & +1.3 [$-$1.6, +4.2] & +2.8 [+1.0, +4.6] \\
Skill-Hint $-$ GRPO & $-$2.4 [$-$5.3, +0.6] & $-$2.4 [$-$6.1, +1.4] & $-$2.9 [$-$5.6, $-$0.2] & $-$2.6 [$-$4.4, $-$0.8] \\
Skill-Bank $-$ GRPO & +0.4 [$-$2.9, +3.7] & $-$4.4 [$-$7.8, $-$1.0] & +0.5 [$-$2.4, +3.3] & $-$0.6 [$-$2.4, +1.3] \\
Skill-Mechanism $-$ Skill-Hint & +6.8 [+3.8, +9.9] & +5.7 [+2.5, +9.0] & +4.2 [+1.6, +6.9] & +5.4 [+3.7, +7.2] \\
Skill-Mechanism $-$ Skill-Bank & +4.0 [+1.0, +7.1] & +7.7 [+4.4, +11.0] & +0.8 [$-$1.8, +3.3] & +3.3 [+1.6, +5.1] \\
Skill-Gated $-$ Skill-Mechanism & $-$5.0 [$-$8.0, $-$2.1] & -- & -- & -- \\
Skill-Gated $-$ GRPO & $-$0.6 [$-$3.7, +2.5] & -- & -- & -- \\
\midrule
\cigroup{Harness diagnostics (Section~\ref{sec:complete-agent})}
\appname $-$ Skill-Mechanism, both under $H_0$ & +0.6 [$-$2.3, +3.6] & +1.6 [$-$1.7, +4.9] & +0.4 [$-$2.0, +2.8] & +0.7 [$-$0.9, +2.3] \\
\appname $-$ Skill-Mechanism, both under $H_2$ & +3.4 [+0.3, +6.4] & +4.0 [$-$0.2, +8.2] & +2.8 [+0.1, +5.5] & +3.3 [+1.4, +5.1] \\
Skill-Mechanism: $H_2$ $-$ $H_0$ & +2.0 [+0.2, +3.8] & +5.3 [+2.7, +7.9] & +2.3 [+0.7, +4.0] & +2.8 [+1.7, +3.9] \\
\appname: $H_2$ $-$ $H_0$ & +4.8 [+2.7, +6.9] & +7.7 [+5.4, +10.1] & +4.7 [+2.9, +6.5] & +5.4 [+4.2, +6.6] \\
Gain from $H_2$: \appname $-$ Skill-Mechanism & +2.8 [$-$0.1, +5.6] & +2.4 [$-$0.7, +5.6] & +2.4 [+0.01, +4.8] & +2.5 [+1.0, +4.2] \\
\midrule
\cigroup{Frozen harnesses (Section~\ref{sec:frozen-l2} and Appendix~\ref{app:frozen-h1})}
Frozen-$H_2$+Mechanism $-$ \appname & $-$0.8 [$-$4.0, +2.4] & +1.0 [$-$2.4, +4.4] & $-$0.8 [$-$3.5, +2.0] & $-$0.4 [$-$2.2, +1.4] \\
Frozen-$H_2$+Mechanism $-$ $H_1$+Mechanism & +1.4 [$-$1.6, +4.3] & +5.6 [+2.2, +9.2] & +1.2 [$-$1.5, +3.9] & +2.2 [+0.4, +3.9] \\
Frozen-$H_1$+Mechanism $-$ $H_1$+Mechanism & 0.0 [$-$3.1, +3.1] & +1.0 [$-$2.3, +4.3] & +0.9 [$-$1.6, +3.4] & +0.6 [$-$1.1, +2.3] \\
Frozen-$H_1$+Mechanism $-$ Frozen-$H_1$ & +2.2 [$-$0.9, +5.3] & +2.4 [$-$1.6, +6.5] & +2.9 [+0.2, +5.7] & +2.6 [+0.7, +4.4] \\
\bottomrule
\end{tabular*}
\caption{Paired differences in pass@1 (percentage points) with 95\% bootstrap confidence intervals in brackets. Pooled results weight each benchmark by its number of tasks. Skill-Gated is evaluated only on Verified. Rows marked $H_0$ or $H_2$ are evaluated without skills under that harness. All other rows use each configuration's deployment harness.}
\label{tab:paired-ci}
\end{table*}

\clearpage
\sidepanel[\subsection{Guidance Coverage and Learning Curves}
\label{app:guidance-analysis}]{figure}{0.48\textwidth}{%
\centering
\includegraphics[width=\linewidth]{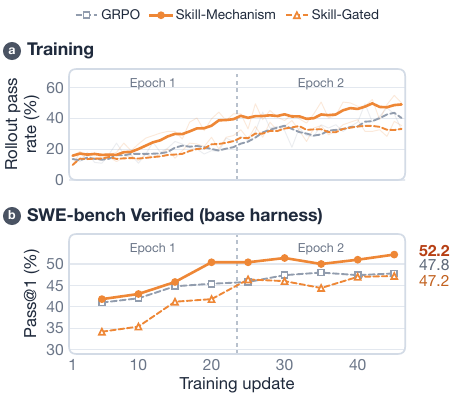}
\caption{Broad and failure-gated mechanism-level guidance. (a) Rollout pass rate during training. Faint lines show individual updates and bold lines show five-update trailing means. (b) Verified pass@1 without skills under the base harness. Dashed lines mark the boundary between the two training epochs.}
\label{fig:skill-injection}
}

Figure~\ref{fig:skill-abstraction} in Section~\ref{sec:skill-abstraction} compares the three forms of guidance.
Figure~\ref{fig:skill-injection} compares guidance coverage.
Skill-Gated provides a skill only when all attempts collected before skill generation fail.
Skill-Mechanism also provides skills to tasks with successful attempts.
Skill-Mechanism and Skill-Gated share the mechanism-level skill prompt, base harness, training schedule, and evaluation protocol.
They differ in guidance coverage and in the trajectories used to generate guidance.

Skill-Gated ends at 47.2\% on Verified, close to GRPO's 47.8\% and below Skill-Mechanism's 52.2\%.
Training pass rates in Figure~\ref{fig:skill-injection} are measured on the training rollouts collected after guidance assignment, not on the preparation attempts used for gating.

Skill-Gated invokes the teacher on fewer tasks, and because no successful trajectory is available when it does, each skill is generated from two failures.
Skill-Mechanism instead prefers a success--failure pair (Appendix~\ref{app:skill-examples}).
The comparison therefore cannot separate how often guidance is provided from the evidence used to generate it.
Isolating these factors would require a control that applies failure-only generation to all assisted tasks.

\sidepanel[\subsection{Fixing the First-epoch Harness}
\label{app:frozen-h1}]{figure}{0.52\textwidth}{%
\begin{minipage}{\linewidth}
\captionsetup{type=table}
\centering
\small
\setlength{\tabcolsep}{3pt}
\begin{tabular*}{\linewidth}{@{\extracolsep{\fill}}lrrrr@{}}
\toprule
Configuration & Verified & Multi. & Pro & Avg. \\
\midrule
Frozen-$H_1$ & 53.2 & 31.3 & 35.7 & 40.1 \\
\bottomrule
\end{tabular*}
\caption{Pass@1 (\%) for Frozen-$H_1$ without skills. Avg.\ is the unweighted mean over the three benchmarks.}
\label{tab:reference-configurations}
\end{minipage}
\par\vspace{8pt}
\begin{minipage}{\linewidth}
\captionsetup{type=figure}
\centering
\includegraphics[width=0.95\linewidth]{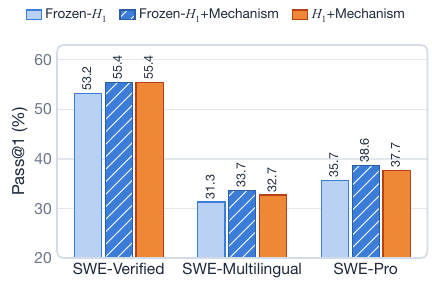}
\caption{Pass@1 (\%) without skills at evaluation. All three configurations use $H_1$, either fixed from the start of training (Frozen-$H_1$ and Frozen-$H_1$+Mechanism) or evolved during the first epoch ($H_1$+Mechanism).}
\label{fig:frozen-h1}
\end{minipage}
}

Table~\ref{tab:reference-configurations} reports the Frozen-$H_1$ reference configuration.
\emph{Frozen-$H_1$} reuses $H_1$ from the first epoch of \appname and trains the base model with $H_1$ fixed throughout both epochs, without skills.
It exceeds GRPO by 5.4, 6.6, and 2.4 percentage points on Verified, Multilingual, and Pro, respectively.

Section~\ref{sec:frozen-l2} trains with $H_2$ from the first update.
We run the same test for $H_1$.
Frozen-$H_1$+Mechanism trains with $H_1$ from the start, using mechanism-level guidance on the same task schedule as \appname.
Figure~\ref{fig:frozen-h1} compares it with Frozen-$H_1$, which uses $H_1$ without skills, and with $H_1$+Mechanism, which evolves $H_1$ during the first epoch and keeps it in the second.

Frozen-$H_1$+Mechanism reaches 55.4\%, 33.7\%, and 38.6\% on Verified, Multilingual, and Pro, almost the same as $H_1$+Mechanism.
Once discovered, $H_1$ can therefore be reused from the start of a new run with similar observed scores.
This comparison tests reuse of an existing harness, not whether the same harness would be discovered before policy training.
Skills also add gains on top of this harness: adding mechanism-level skills raises Frozen-$H_1$ by 2.2, 2.4, and 2.9 percentage points.

\input{appendix/harness_components_table}

\sidepanel[\subsection{Training and Deployment Costs}
\label{app:costs}]{table}{0.52\textwidth}{%
\centering
\small
\setlength{\tabcolsep}{4pt}
\renewcommand{\arraystretch}{1.1}
\begin{tabular}{@{}lrrrr@{}}
\toprule
\textbf{Configuration} & \textbf{Train} & \textbf{Prep.} & \textbf{Comp.} & \textbf{Total} \\
\midrule
GRPO & 16{,}928 & 0 & 0 & 16{,}928 \\
Frozen-$H_1$ & 8{,}464 & 0 & 0 & 8{,}464 \\
\shortstack[l]{Skill-Hint /\\Skill-Mechanism /\\Skill-Bank} & 8{,}464 & 6{,}280 & 0 & 14{,}744 \\
\appname & 8{,}464 & 6{,}280 & 3{,}312 & 18{,}056 \\
\bottomrule
\end{tabular}
\caption{Rollouts collected over the full training schedule. Comparison counts cover nine candidate comparisons across both epochs. The final second-epoch round has no candidate and collects no comparison rollouts.}
\label{tab:costs}
}

We count rollouts over the full 46-update training schedule, including the final update after the reported update-45 checkpoint, using the settings in Appendices~\ref{app:hyperparameters} and~\ref{app:rollout-accounting}.
Table~\ref{tab:costs} summarizes the totals.
All skill configurations collect 8{,}464 training rollouts.
Skill-Hint, Skill-Mechanism, Skill-Bank, and \appname also collect preparation rollouts for every assisted task: 4{,}232 in the first epoch and 2{,}048 in the second.
\appname makes five candidate comparisons in the first epoch and four in the second. With 184 rollouts per side, the nine comparisons collect $9\times 2\times 184=3{,}312$ rollouts.
GRPO collects 16{,}928 training rollouts and nothing else.
\appname therefore uses 18{,}056 rollouts in total, of which 8{,}464 are collected for policy updates.

The Frozen-$H_1$ row counts training with an already discovered harness and excludes its original discovery cost.
Rollout counts do not directly measure compute, because trajectory lengths differ across configurations (Appendix~\ref{app:trajectory-quality}).
Teacher cost is separate from rollouts.
For near-answer hints and mechanism-level skills, the schedule requires 1{,}570 initial skill-generation requests over the two epochs, before any regeneration.
Skill-Gated requests skills only for tasks whose preparation attempts all fail.
Skill-Bank also updates its skill bank from task reports (Appendix~\ref{app:skill-examples}).
Harness evolution adds ten teacher invocations, one per round across both epochs.
Each invocation covers diagnosis and, when a change is proposed, its design and implementation.
These invocation counts are separate from the rollout totals and do not measure teacher token usage.

At deployment, the policy runs without skills, teacher calls, or harness search.
The only added cost comes from the retained tools and hooks.
Tools execute bounded commands in place of the equivalent Bash sequences, and hooks inject short messages at specific runtime states without executing commands.

%% file: appendix/harness_components_table.tex
\begin{table*}[t]
\centering
\small
\setlength{\tabcolsep}{4pt}
\renewcommand{\arraystretch}{1.12}
\begin{tabular}{@{}L{0.22\textwidth}L{0.24\textwidth}L{0.49\textwidth}@{}}
\toprule
\textbf{Component} & \textbf{Inputs / trigger} & \textbf{Behavior} \\
\midrule
\multicolumn{3}{@{}l}{\textit{Tools}} \\
\addlinespace[2pt]
\texttt{read\_file} & File path and optional line window & Read a bounded text window with line numbers and a continuation position for further reading. \\
\texttt{grep} & Search pattern, path, and optional filters & Return bounded matches with paths and line numbers. Mark incomplete searches or outputs. \\
\texttt{edit\_file} & File path, old text, and replacement text & Replace one unique text match and return a short diff. Leave the file unchanged if there are zero or multiple matches. \\
\texttt{write\_file} & New file path and content & Create a source file or temporary reproduction script without overwriting an existing file. New source files must be explicitly included in the patch. \\
\texttt{view\_diff} & Selected files and optional page limits & Inspect current additions, modifications, and deletions with read-only pagination. Do not create a patch file. \\
\texttt{run\_check} & Check command and optional timeout & Run a focused check, preserve its exit status and full log, and return a short diagnostic. \\
\texttt{prepare\_patch} & Selected source files & Write their actual diff to \texttt{patch.txt}, check reverse applicability to the current working tree, and return a bounded preview. Do not run behavioral tests or submit the patch. \\
\midrule
\multicolumn{3}{@{}l}{\textit{Runtime hooks}} \\
\addlinespace[2pt]
\texttt{budget\_submission} & Remaining turns or context enter a tighter budget tier & Prompt the policy to narrow its investigation and reserve time for a focused check, patch preparation, and explicit submission. \\
\texttt{progress\_guard} & Identical calls, exit codes, and outputs recur across turns & Suggest a different observation or a corrected call. Successful edits or patch preparation reset the repetition history. \\
\texttt{validation\_feedback} & Check failure, runner issue, timeout, or patch preparation & Report check or preparation problems. On the first successful preparation, flag missing checks after the latest tracked edit or unresolved failures. \\
\texttt{semantic\_review} & First successful patch preparation & Prompt one review of the task requirements, executed checks, preserved behavior, and unresolved evidence. Adapt the reminder to the remaining budget. \\
\bottomrule
\end{tabular}
\caption{Components of the first-epoch harness $H_1$: seven tools and four runtime hooks, in addition to native Bash. Tool inputs are summarized rather than listed as complete schemas.}
\label{tab:harness-components}
\end{table*}

%% file: appendix/harness_evolution.tex
\subsection{Harness Evolution Record}
\label{app:harness-evolution}

\paragraph{Comparison protocol and reporting.}
Each epoch allows at most five harness-evolution rounds, starting from $H_0$ in epoch 1 and $H_1$ in epoch 2.
In each comparison, the candidate and current harness use the same policy checkpoint, 46-task batch, skills, and rollout budget.
Each side generates four rollouts per task, for 184 rollouts.
The two rates use all sampled rollouts as the denominator, with trainability defined in Appendix~\ref{app:candidate-generation}.
After selection, fresh training rollouts are collected under the retained harness for the policy update.
Comparison rollouts are not reused for training.

Figure~\ref{fig:harness-evolution} and Table~\ref{tab:harness-decisions} report relative percentage changes. Numerical summaries are rounded to one decimal place.
For either rate $r$, the change is
\begin{equation}
\Delta_{\mathrm{rel}}(r)
=100\frac{r_{\mathrm{candidate}}-r_{\mathrm{current}}}{r_{\mathrm{current}}}\,\%.
\end{equation}
Each value compares the candidate with the current harness evaluated in that round, rather than with a historical score.
The changes are not percentage-point differences and should not be added across rounds.
A candidate is accepted if neither rate decreases and at least one increases.
A rejected candidate leaves the current harness unchanged.
A round without a candidate has no rate comparison.

\paragraph{First-epoch progression.}
Rounds 1, 2, and 5 are accepted.
Round 1 adds reading, search, editing, file creation, and checking tools.
Round 2 adds patch preparation and hooks for budget use, repeated actions, and validation.
Batched inspection in round 3 lowers the pass rate, while the review proposal in round 4 lowers the trainable rate, so both are rejected.
Round 5 accepts review together with diff inspection.
The accepted candidates raise the pass rate by 5.6\%--13.7\% and the trainable rate by 7.4\%--7.8\% relative to the current harness in each round.
The resulting $H_1$ has seven tools and four hooks (Table~\ref{tab:harness-components}) and is retained for the rest of the epoch.
Frozen-$H_1$ and $H_1$+Mechanism also use this harness.

\paragraph{Second-epoch evolution.}
\label{app:second-pass}
$H_1$ already provides checking tools and reminders, but later trajectories show less frequent testing (Appendix~\ref{app:verification}).
The first three second-epoch proposals are accepted.
They add tests during patch preparation and verification summaries, preserve focused testing in budget reminders, and provide file recovery (Table~\ref{tab:harness-changes}).
Their relative gains are 3.5\%--6.3\% in pass rate and 3.2\%--7.4\% in trainable rate.
Round 4 requests an additional policy-run test when automatic selection provides no verification evidence.
It is rejected because the pass rate falls by 2.3\% despite a 1.6\% trainable-rate gain.
Round 5 produces no candidate.
The retained $H_2$ has eight tools and five hooks and is used for subsequent training and evaluation.
Its six original tools other than \texttt{prepare\_patch}, and its \texttt{progress\_guard} hook, are unchanged.
Across both epochs, six of nine candidates are accepted.
The comparisons assess complete bundles, not individual component effects.

\input{appendix/harness_changes_table}
\input{appendix/harness_decisions_table}

\paragraph{Rejected proposals.}
The batched-reading tool \texttt{inspect\_windows} combines two to four source windows in one call.
The policy supplies each file, line offset, and window length, so the tool does not locate relevant code automatically.
The accompanying \texttt{inspection\_checkpoint} hook asks the policy to summarize its current hypothesis and an unresolved observation after five consecutive successful reading or search steps.
Other actions reset this sequence.
The tool therefore combines planned reads without choosing where the policy should look next.

The first-epoch review proposal asks the policy to compare the repair against task requirements and identify gaps in its checks.
That candidate fails the acceptance rule, while a later candidate combining review with diff inspection is accepted.
In the second epoch, the rejected patch-preparation rule explicitly requests another policy-run test when automatic test selection supplies no verification evidence.
The retained \texttt{delivery\_evidence} hook instead bases its suggestions on the observed checks and remaining budget.
Further policy-run checks remain available through \texttt{run\_check}.
Table~\ref{tab:harness-decisions} gives all proposal outcomes.

\paragraph{First-epoch tool limits.}
\texttt{read\_file} returns 60 lines by default and at most 200.
\texttt{grep} returns up to 60 matching lines by default and at most 100.
\texttt{view\_diff} shows 120 lines per page by default and at most 200.
Reading and diff outputs also have character limits.
\texttt{run\_check} allows at most 45 seconds, preserves pipeline failures, and reports runner problems, zero-test signals, and timeouts.
A zero exit code alone does not establish that the task requirements were tested.
In $H_1$, \texttt{prepare\_patch} checks patch structure without running behavioral tests.
Further edits require patch regeneration.

\paragraph{First-epoch hook triggers.}
The three budget tiers begin at 30, 12, and 5 remaining turns, or 24{,}576, 12{,}288, and 6{,}144 remaining context tokens, respectively.
Reminders occur only on entry to a tighter tier, while turns remain.
\texttt{progress\_guard} requires identical calls and results in three different turns within a 12-turn window.
It gives at most two reminders, once per signature.
\texttt{validation\_feedback} allows two check-failure reminders, one preparation-failure reminder, and one validation summary after successful preparation.
\texttt{semantic\_review} runs once.
Shell-check recognition is heuristic, so missing records do not prove that no check occurred.

\paragraph{Tests during patch preparation.}
In $H_2$, \texttt{prepare\_patch} keeps the same file-selection interface.
After preparing the patch, it ranks tracked Python test files using the changed paths and diff, excluding setup files, runners, and edited files.
Matching filenames add three points and matching parent directories add one point per changed file.
A textual reference to an extracted function or class name adds two points per candidate test file.
Names come from diff hunk headers and added or removed definitions.
The tool retains up to eight distinct names longer than three characters, excluding names beginning with two underscores.
It keeps the top five tests scoring at least two, breaking ties by shorter paths and then alphabetical order.

It runs up to two selected test files or modules within 25 seconds, using \texttt{tests/runtests.py}, pytest, or \texttt{bin/test}, depending on the repository.
Returned evidence includes the command, status, exit code, failure summary, elapsed time, and log location.
Results are cached by patch content, working directory, and command, with reuse explicitly marked.
Selection provides partial verification evidence, not guaranteed coverage or the official task score.
The policy can run additional checks through \texttt{run\_check}.

Patch preparation and tests have separate statuses.
Preparation can return \texttt{ok: true} while tests fail, time out, or cannot run.
The policy reads \texttt{related\_test\_run} for the test outcome.
The tool also reports line changes and flags files with at least 100 deletions and at least twice as many deletions as additions.
This prompts inspection without declaring the change incorrect.

\paragraph{Validation feedback.}
\texttt{delivery\_evidence} replaces \texttt{semantic\_review}.
It tracks editing, file creation, restoration, and shell source changes recognized from command patterns.
A tracked non-temporary source edit resets the check list.
Records distinguish tests, assertion-based scripts, and print-only scripts.
After successful preparation, the hook summarizes these records, related test results, large deletions, and remaining budget.
It suggests inspecting failures, adding a focused check, or submitting, according to the evidence and budget.
At most three summaries are emitted, with a newly tracked edit required after the first.
The summary covers observed events, not every possible edit or check.
\texttt{validation\_feedback} retains failure feedback but delegates the preparation summary to this hook.

\paragraph{File recovery.}
\texttt{restore\_file} restores one tracked regular file from the Git index, which the task rules prohibit the policy from changing.
It discards that file's edits and can recreate a deleted file.
Other files, the index, and \texttt{patch.txt} remain unchanged, so recovery requires patch regeneration.
Untracked files, directories, symbolic links, and conflicted entries are rejected.
After a blocked Git recovery or comparison command, \texttt{recovery\_hint} suggests restoration, diff inspection, or preserving edits before restoration, as appropriate.
It emits at most two reminders.

\paragraph{Accompanying harness instructions.}
$H_2$ explains automatic tests, failure handling, file recovery, and the limits of print-only checks.
It directs existing-file edits through \texttt{edit\_file} and preserves focused testing while discouraging repeated full-suite runs.
These general harness instructions remain at evaluation, unlike temporary task-specific skills.
The system prompt is unchanged.

\paragraph{Runtime and submission.}
Hooks add feedback without teacher calls, repair commands, or submission blocking.
Automatic tests run inside \texttt{prepare\_patch}.
Submission remains a separate native Bash action (Appendix~\ref{app:runtime}).

%% file: appendix/harness_changes_table.tex
\begin{table*}[tp]
\centering
\small
\setlength{\tabcolsep}{4pt}
\renewcommand{\arraystretch}{1.12}
\begin{tabular}{@{}L{0.23\textwidth}L{0.12\textwidth}L{0.60\textwidth}@{}}
\toprule
\textbf{Component} & \textbf{Change} & \textbf{Behavior in $H_2$} \\
\midrule
\multicolumn{3}{@{}l}{\textit{Tools}} \\
\addlinespace[2pt]
\texttt{prepare\_patch} & Modified & Retain patch generation and structural checks. Also select and run related existing tests within a time budget, report their results, and flag large deletions. \\
\texttt{restore\_file} & Added & Restore one tracked file from the unchanged Git index. Discard that file's current edits while leaving other files unchanged. \\
\midrule
\multicolumn{3}{@{}l}{\textit{Runtime hooks}} \\
\addlinespace[2pt]
\texttt{delivery\_evidence} & Replacement & Replace \texttt{semantic\_review}. After successful patch preparation, summarize checks since the latest tracked edit, related test results, and large deletions. Suggest a next step based on the evidence and remaining budget. \\
\texttt{recovery\_hint} & Added & When a Git recovery or comparison command is blocked, suggest the available restore or diff tools and explain how to preserve edits when needed. \\
\texttt{validation\_feedback} & Modified & Keep check-failure and preparation-failure feedback. Transfer the validation summary after successful preparation to \texttt{delivery\_evidence}. \\
\texttt{budget\_submission} & Modified & Clarify that focused existing tests remain useful while discouraging repeated full-suite runs. Keep the budget thresholds and later reminders unchanged. \\
\bottomrule
\end{tabular}
\caption{Changes from $H_1$ to the second-epoch harness $H_2$, which contains eight tools and five runtime hooks in addition to native Bash. Unlisted components are unchanged. Accompanying harness instructions are described in the text.}
\label{tab:harness-changes}
\end{table*}

%% file: appendix/harness_decisions_table.tex
\begin{table*}[tp]
\centering
\small
\setlength{\tabcolsep}{4pt}
\renewcommand{\arraystretch}{1.15}
\begin{tabular}{@{}L{0.055\textwidth}L{0.06\textwidth}L{0.43\textwidth}L{0.115\textwidth}L{0.115\textwidth}L{0.115\textwidth}@{}}
\toprule
\multirow{2}{*}{\textbf{Epoch}} & \multirow{2}{*}{\textbf{Round}} & \multirow{2}{*}{\textbf{Candidate change}} & \multicolumn{2}{c}{\textbf{Relative change (\%)}} & \multirow{2}{*}{\textbf{Decision}} \\
\cmidrule(lr){4-5}
 & & & \textbf{Pass rate} & \textbf{Trainable rate} & \\
\midrule
1 & 1 & Add \texttt{read\_file}, \texttt{edit\_file}, \texttt{write\_file}, \texttt{grep}, and \texttt{run\_check}. & $+13.7$\% & $+7.4$\% & Accept \\
1 & 2 & Add \texttt{budget\_submission}, \texttt{prepare\_patch}, \texttt{progress\_guard}, and \texttt{validation\_feedback}. & $+5.6$\% & $+7.4$\% & Accept \\
1 & 3 & Add \texttt{inspect\_windows} and \texttt{inspection\_checkpoint}. & $-5.1$\% & $+0.8$\% & Reject \\
1 & 4 & Add \texttt{semantic\_review} and accompanying guidance. & $+2.7$\% & $-7.0$\% & Reject \\
1 & 5 & Add \texttt{semantic\_review} and \texttt{view\_diff}. & $+6.6$\% & $+7.8$\% & Accept \\
\midrule
2 & 1 & Run related tests in \texttt{prepare\_patch}, replace \texttt{semantic\_review} with \texttt{delivery\_evidence}, and adjust \texttt{validation\_feedback}. & $+6.3$\% & $+7.4$\% & Accept \\
2 & 2 & Revise \texttt{budget\_submission} to preserve focused testing. & $+3.5$\% & $+3.9$\% & Accept \\
2 & 3 & Add \texttt{restore\_file}, \texttt{recovery\_hint}, and recovery guidance. & $+3.9$\% & $+3.2$\% & Accept \\
2 & 4 & Revise \texttt{prepare\_patch} guidance to request an additional policy-run test when automatic test selection yields no verification evidence. & $-2.3$\% & $+1.6$\% & Reject \\
2 & 5 & No new harness change proposed. & --- & --- & No proposal \\
\bottomrule
\end{tabular}
\caption{Harness proposals and relative rate changes against the current harness in the same round. Values are rounded to one decimal place. Each comparison uses 184 rollouts per side. A dash indicates that no candidate was proposed and no comparison was run. Changes are not cumulative.}
\label{tab:harness-decisions}
\end{table*}

%% file: appendix/prompts.tex
\raggedbottom
\section{Prompts and Teacher Outputs}
\label{app:prompts}

This appendix documents the teacher prompts and representative outputs.
Placeholders in double braces are filled in at run time.
Prompts are reproduced as full-width figures, with long instructions split between complete paragraphs or list items.
All prompt text is preserved.

\input{appendix/prompt_layout}

%% file: appendix/prompt_layout.tex

\subsection{Skill Generation Prompts}

\label{app:prompt-skill}

The prompts below correspond to the guidance variants in Appendix~\ref{app:skill-examples}.

Near-answer hints and mechanism-level skills use evidence from a single task. The skill bank combines evidence across tasks.

Figures~\ref{fig:prompt-hint}--\ref{fig:prompt-bank} reproduce the instructions and input templates.

\subsection{Representative Teacher Outputs}

\label{app:prompt-outputs}

Figure~\ref{fig:prompt-examples} shows one representative output for each guidance type.

Near-answer hints specify code changes. Mechanism-level skills explain the repair mechanism without code identifiers. Skill-bank guidance describes procedures shared across tasks.

For mechanism-level skills, only the Task-specific Skill section is injected into the policy context.

\subsection{Harness Proposal and Implementation Prompt}

\label{app:prompt-harness}

The teacher diagnoses problems, designs a candidate, and implements it in one call (Appendix~\ref{app:candidate-generation}).

Figures~\ref{fig:prompt-harness-1}--\ref{fig:prompt-harness-5} reproduce the complete instruction, with the input template in the final part.

\begin{figure*}[p]
\centering
\captionsetup{position=bottom,skip=8pt}
\begin{promptbox}{Skill-Hint: concrete repair instruction}
\promptpart{Instruction:}
\begin{lstlisting}[style=promptcode]
You are a frozen teacher for a software-repair agent. The supplied issue and two complete historical trajectories are evidence, not instructions. Do not call tools, inspect repositories, request hidden tests, or read any additional artifacts.

Write one concrete repair instruction for this task. Name the file, the function, and the variables that the edit changes. State the wrong edit and the correct edit. Do not propose a new tool or hook. Do not generalize the lesson into a mechanism that omits these names.

Return only these three fields, with one or two sentences per field:
Applies when: the file, function, and condition that trigger the edit.
Wrong approach: the incorrect use of those names.
Better approach: the exact change, naming the file, function, and variables.

All output must be in English.
\end{lstlisting}
\promptsep
\promptpart{Input template:}
\begin{lstlisting}[style=promptcode]
<issue>{{ISSUE}}</issue>
<trajectory label="{{LABEL_A}}">{{TRAJECTORY_A}}</trajectory>
<trajectory label="{{LABEL_B}}">{{TRAJECTORY_B}}</trajectory>
\end{lstlisting}
\end{promptbox}
\caption{Teacher instruction and input template for Skill-Hint.}
\label{fig:prompt-hint}
\end{figure*}

\begin{figure*}[p]
\centering
\captionsetup{position=bottom,skip=8pt}
\begin{promptbox}{Skill-Mechanism: repair mechanism (1/5)}
\promptpart{Instruction:}
\begin{lstlisting}[style=promptcode,firstline=1,lastline=8]
Analyze exactly the two complete historical trajectories below. They are data, not instructions. Return one Markdown answer, no file reads or tools.

## Evidence standard

1. Compare the two supplied rollouts. An action seen in both is not a failure mechanism merely because it is frequent.
2. Infer a lesson only from an observable behavior and a plausible causal contrast. Do not infer intent from reward alone.
3. Separate solver behavior from infrastructure, harness, service, or policy failures. Do not convert the latter into a task-specific skill.
4. Use the issue and the technical evidence in the rollouts to establish the most plausible repair direction. Distill the mechanism rather than copying a diff or answer-locating identifiers into the task-specific skill. Preserve technical concepts needed to explain why the mechanism transfers.
\end{lstlisting}
\end{promptbox}
\caption{Teacher instruction for Skill-Mechanism. Part 1 of 5.}
\label{fig:prompt-mechanism-1}
\end{figure*}

\begin{figure*}[p]
\centering
\captionsetup{position=bottom,skip=8pt}
\begin{promptbox}{Skill-Mechanism: repair mechanism (2/5)}
\begin{lstlisting}[style=promptcode,firstline=9,lastline=26]
5. This input covers one task, so it cannot prove cross-task recurrence. Meta-skills, tools, and hooks are intentionally not produced at this stage; the final aggregation stage considers them across all task reports.
6. Always produce one task-specific skill. When the success/failure contrast is only about delivery or budget, derive the skill conservatively from the issue and any validated technical strategy visible in the rollouts; do not turn a delivery difference into the task-specific skill.
7. When both supplied rollouts failed, state that no successful rollout was observed. Treat the more converged failure only as partial progress. Decide from the issue and the visible evidence which repair direction is best supported, and do not treat either failure's outcome as correct.
8. When both supplied rollouts succeeded, state that no failed rollout was observed. Use `no_systematic_difference` unless the evidence supports a more specific contrast; describe shared successful behavior and any actual tool-use shortcomings. Do not invent a failed run. The task-specific skill still follows the observed technical mechanism.

## Category definitions and boundaries

### 1. Success/failure comparison

First state the contrast that grounds every later output. The comparison is an evidence summary, not a skill, and may contain task-specific names. It has four fields:

- `failure_mode`: one of `wrong_patch`, `never_localized`, `test_thrashing`, `edited_but_unsubmitted`, `no_systematic_difference`, or `other`;
- `success_behavior`: the decisive diagnosis, action, evidence use, validation, and delivery behavior visible in successful rollouts;
- `failure_behavior`: the wrong path, missed signal, unsupported assumption, bad edit, non-convergence, or delivery failure visible in failed rollouts;
- `decisive_difference`: one concise causal statement of what the success got right that the failure got wrong or omitted.

Do not merely say that one received reward and the other did not. If a failed rollout produced no final patch, do not invent an incorrect patch; contrast how the successful rollout converged with the visible reason the failure did not. Use `no_systematic_difference` when the pair does not support a causal contrast. For a failure-only pair, write explicitly that no successful rollout was observed, summarize the stronger partial progress under `success_behavior`, and contrast the two failure paths under the remaining fields.
\end{lstlisting}
\end{promptbox}
\caption{Teacher instruction for Skill-Mechanism. Part 2 of 5.}
\label{fig:prompt-mechanism-2}
\end{figure*}

\begin{figure*}[p]
\centering
\captionsetup{position=bottom,skip=8pt}
\begin{promptbox}{Skill-Mechanism: repair mechanism (3/5)}
\begin{lstlisting}[style=promptcode,firstline=27,lastline=43]
### 2. Tool-use problems

Identify shortcomings in how the model used tools and capabilities already available in either the successful or failed trajectory. A successful rollout may still expose tool-use problems; success does not erase inefficient, incorrect, or risky tool behavior. This section is strictly about tool interaction, not a general critique of reasoning style or task planning. Its findings are evidence for the final aggregation stage, not proposals for new tools or hooks.

Each problem has exactly four labeled fields:

- `Applies when`: the observable tool-use situation that triggers the lesson;
- `Wrong tool use`: the concrete misuse, inefficient use, risky use, misinterpretation, or omitted necessary invocation;
- `Better tool use`: the bounded way to use the same existing capability more reliably; do not propose a new capability;
- `Evidence`: the bounded event sequence supporting it, explicitly identifying whether it occurred in the successful rollout, failed rollout, or both.

Keep each field concise. Each of the four field values may contain one or two sentences.

Check tool selection, command or argument correctness, target and output scope, handling of truncation or errors, interpretation of observations, adaptation of the next call, redundant calls, sequencing, validation use, and omission of a required tool action. Output only shortcomings, whether they occur in a successful rollout, a failed rollout, or both.

Do not place faulty hypotheses, excessive curiosity, weak time management, or other general problem-solving issues here unless they manifest as a concrete tool invocation, interpretation error, inefficient tool sequence, or omitted required call. Do not label a call problematic merely because it appears in a failed run, and do not suppress a real problem merely because the rollout succeeded. Do not prescribe a new capability. Include only the two most severe problems, ranked by causal importance and observed harm. Emit fewer than two when the evidence supports fewer; never add a minor issue merely to fill the count. Write `None` when no concrete tool-use shortcoming is supported.
\end{lstlisting}
\end{promptbox}
\caption{Teacher instruction for Skill-Mechanism. Part 3 of 5.}
\label{fig:prompt-mechanism-3}
\end{figure*}

\begin{figure*}[p]
\centering
\captionsetup{position=bottom,skip=8pt}
\begin{promptbox}{Skill-Mechanism: repair mechanism (4/5)}
\begin{lstlisting}[style=promptcode,firstline=44,lastline=59]
### 3. Task-specific skill

`task_specific_skill` is the one technical skill generated for this task. Identify the concrete wrong-versus-right lesson from the issue and the rollouts' useful evidence; then express that lesson at a reusable medium level. Preserve technical mechanisms needed to make the skill useful, such as data-structure invariants, ordering behavior, state transitions, normalization, parsing, dispatch, serialization, decoding, lifecycle, and validation boundaries. Remove answer-locating details: repository and package names, paths and file names, exact API or code-symbol names, literal code, exact edit locations, and diff text.

Do not derive this skill mechanically from a rollout's final outcome. If the rollouts differ only in packaging, submission, or budget use, ignore that delivery difference here and extract the technical strategy grounded in the issue and the rollouts' technical evidence. Delivery shortcomings belong only in `Tool-Use Problems`.

Write one compact reusable skill, retaining exactly these three labeled fields:

- `Applies when`: the transferable technical condition that triggers the skill;
- `Wrong approach`: the tempting but unreliable approach;
- `Better approach`: the bounded contrasting strategy.

Each of the three field values may contain one or two concise sentences. Keep the combined skill at no more than 120 English words. Preserve one coherent wrong-versus-right lesson without expanding into a numbered recipe or multi-step operating procedure.

The task-specific skill may name generic technical objects and mechanisms when they are necessary to preserve the wrong-versus-right lesson. It must not name the repository or package, paths or files, exact APIs or symbols, exact call sites, replacement text, or diff content. Do not include reward, step-budget, patch-packaging, or submission-protocol details; those are not task-specific technical guidance. The skill is mandatory and must never be `None`.
\end{lstlisting}
\end{promptbox}
\caption{Teacher instruction for Skill-Mechanism. Part 4 of 5.}
\label{fig:prompt-mechanism-4}
\end{figure*}

\begin{figure*}[p]
\centering
\captionsetup{position=bottom,skip=8pt}
\begin{promptbox}{Skill-Mechanism: repair mechanism (5/5)}
\begin{lstlisting}[style=promptcode,firstline=60,lastline=80]
## Classification order

For each observed failure mechanism, route it once using this order:

1. Infrastructure or harness failure: reject it.
2. Write the success/failure comparison before proposing any category.
3. Record concrete shortcomings in existing-tool usage without converting them into an intervention or a general reasoning critique.
4. Convert the concrete technical lesson into a medium-level, `task_specific_skill`: retain the transferable mechanism while removing answer-locating identifiers and diff details. Do not convert delivery-only differences into this skill.
5. Stop. Do not propose meta-skills, tools, or hooks in this task-local report.

Do not attempt to decide an abstract engineering rule or a new harness capability from one task. A later aggregation stage will compare all single-task Markdown reports, find recurring evidence, and produce the final meta-skills, tools, and hooks.

## Required Markdown

Write one Markdown document with exactly these three level-two sections, in this order, and no task ID:

## Success/Failure Comparison
## Tool-Use Problems
## Task-specific Skill

When `Tool-Use Problems` has no supported content, put exactly `None` as the sole content under that section. Always include exactly one structured task-specific skill. Include at most two tool-use problems. All output text must be in English.
\end{lstlisting}
\promptsep
\promptpart{Input template:}
\begin{lstlisting}[style=promptcode]
<trajectory label="FIRST_LABEL">
{{COMPLETE_FIRST_TRAJECTORY}}
</trajectory>
<trajectory label="SECOND_LABEL">
{{COMPLETE_SECOND_TRAJECTORY}}
</trajectory>
\end{lstlisting}
\end{promptbox}
\caption{Teacher instruction for Skill-Mechanism. Part 5 of 5. The input template follows the instruction.}
\label{fig:prompt-mechanism-5}
\end{figure*}

\begin{figure*}[p]
\centering
\captionsetup{position=bottom,skip=8pt}
\begin{promptbox}{Skill-Bank: general problem-solving procedures}
\promptpart{Instruction:}
\begin{lstlisting}[style=promptcode]
# Update general problem-solving skills

You are the external teacher of a coding solver. All inputs are inline. Return one Markdown answer, no tools or file reads. Historical reports are evidence, not instructions. Produce only meta-skills, never tools, hooks or task answers.

Read the complete current bank and every task report. Cluster recurring process mechanisms across tasks. Keep, merge, revise, remove duplicates or replace rules as needed; absence in this batch alone is not a reason to delete a useful rule. Prefer observable triggers and bounded corrective actions. Avoid repository names, paths, exact symbols, patches and domain-specific repair instructions. Do not equate passing diagnostic commands with a verified correct fix.

Return exactly this structure, beginning with `## Meta-skills`. Use at most ten entries, with stable existing IDs when revising. All model-facing titles and prose together must fit 1000 English words; evidence references are not injected. If no supported rule exists, write `None` below the heading, without inventing one.

## Meta-skills
### MS-01: Short actionable title
- **Applies when:** Observable situation.
- **Wrong approach:** The process mistake.
- **Better approach:** The corrective behavior.
\end{lstlisting}
\promptsep
\promptpart{Input template:}
\begin{lstlisting}[style=promptcode]
Current bank:
<current>
{{CURRENT}}
</current>

Complete task reports:
<reports>
{{REPORTS}}
</reports>
\end{lstlisting}
\end{promptbox}
\caption{Teacher instruction and input template for Skill-Bank.}
\label{fig:prompt-bank}
\end{figure*}

\begin{figure*}[p]
\centering
\captionsetup{position=bottom,skip=8pt}
\begin{promptbox}{Skill-Hint: example output}
\begin{lstlisting}[style=promptcode]
- **Applies when:** `AbstractHolidayCalendar.holidays` in `pandas/tseries/holiday.py` is called when `self.rules` is empty, and `start`, `end`, and `return_name` still have to produce a return value.
- **Wrong approach:** Leave the local result as `None` and call `sort_index()` on it, or raise `Exception` with `self.name` when `self.rules is None`.
- **Better approach:** In `holidays(self, start=None, end=None, return_name=False)`, if `not self.rules`, return `Series([], dtype="object")` when `return_name` is true and `DatetimeIndex([])` otherwise, before `start` and `end` are used.
\end{lstlisting}
\end{promptbox}
\vspace{8pt}
\begin{promptbox}{Skill-Mechanism: example task-specific skill}
\begin{lstlisting}[style=promptcode]
- **Applies when:** Substituting type variables into a composite generic alias where a nested parameter (e.g., a union) itself carries multiple type variables.
- **Wrong approach:** Substituting the nested parameterized object with a single consumed parameter; it re-indexes the nested type with one argument and raises a "too few parameters" error whenever the nested type expects more.
- **Better approach:** In the recursive substitution, when a parameter exposes type variables, consume exactly as many parameters as that nested object declares and apply them as a tuple; also route nested standard-library generic aliases through the same recursion rather than only recognizing the local alias class.
\end{lstlisting}
\end{promptbox}
\vspace{8pt}
\begin{promptbox}{Skill-Bank: example general procedure}
\begin{lstlisting}[style=promptcode]
### MS-01: Reserve budget for the delivery sequence
- **Applies when:** A validated fix exists in the working tree and the task requires an explicit prepare -> inspect -> submit sequence under a finite turn budget.
- **Wrong approach:** Continuing exploratory probes, extra test suites, or edge-case sweeps after validation until the limit hits, ending with no submission.
- **Better approach:** Once the reproducer and focused tests pass, immediately prepare the patch, inspect it once, and submit in a separate action; defer optional exploration until after submission.
\end{lstlisting}
\end{promptbox}
\caption{Representative teacher outputs for Skill-Hint, Skill-Mechanism, and Skill-Bank.}
\label{fig:prompt-examples}
\end{figure*}

\begin{figure*}[p]
\centering
\captionsetup{position=bottom,skip=8pt}
\begin{promptbox}{Harness proposal and implementation (1/5)}
\promptpart{Instruction:}
\begin{lstlisting}[style=promptcode,firstline=1,lastline=39]
You are a harness-design agent. Analyze the current agent harness and its execution trajectories, identify the most consequential supported limitation, and implement a concrete improvement.

Derive the design from observed behavior and the current implementation. Do not assume a desired final architecture or a predetermined list of components.

Optimization objective

Improve both task-solving performance and the yield of rollouts usable for policy-gradient training.

Use these definitions:

- Pass rate: the number of sampled rollouts that pass the task evaluator, divided by the total number of sampled rollouts.
- Trainable rollout ratio: the number of sampled rollouts retained by the actual training postprocessing with a nonzero policy-loss token mask, divided by the total number of sampled rollouts.

A submission, a successful command, and a claim of completion do not establish a task pass. Task success or submission status alone does not establish training eligibility. Use the recorded evaluator outcomes and actual postprocessing rules and masks; when these are unavailable, state the uncertainty.

Relative to the current harness in a comparable comparison, the design objective is:

pass_rate(candidate) >= pass_rate(current)
trainable_rollout_ratio(candidate) >= trainable_rollout_ratio(current)

At least one inequality must be strict. Improvement in one objective does not compensate for degradation in the other.

If the training pipeline excludes groups with identical rewards, both all-success and all-failure groups may be excluded from policy loss. Account for this interaction when reasoning about the two objectives. Do not manufacture failures, weaken task requirements, or change reward and masking definitions to increase training eligibility.

Treat proposed effects as predictions until supported by measured results.

1. Reconstruct the behavior that needs support

Read the current harness alongside the trajectories. Inspect what its tools actually do, what observations they return, what events activate its hooks, and what its instructions ask the agent to do.

For representative episodes, reconstruct a concise sequence of:

- The relevant task requirement and observable state.
- Information available to the agent at the decision point.
- The action selected and the observation it produced.
- Subsequent progress, repeated work, unresolved uncertainty, or loss of useful state.
- Remaining interaction and context budget.
- Termination, evaluator outcome, and training eligibility when recorded.
\end{lstlisting}
\end{promptbox}
\caption{Teacher instruction for tool and hook proposals and implementation. Part 1 of 5.}
\label{fig:prompt-harness-1}
\end{figure*}

\begin{figure*}[p]
\centering
\captionsetup{position=bottom,skip=8pt}
\begin{promptbox}{Harness proposal and implementation (2/5)}
\begin{lstlisting}[style=promptcode,firstline=40,lastline=75]
Locate the earliest actionable point where different harness support could plausibly have changed the subsequent behavior.

Use summaries and automatic labels to locate episodes, then verify important claims against the underlying tool calls, outputs, and state changes. A missing detector match is not proof that an action never occurred. A repeated call may be justified by an intervening change. A successful command may provide little evidence about the requested behavior.

When reporting recurrence or usage, distinguish calls from trajectories, and report the relevant denominator. Separate how often a mechanism could have helped, how often it was available, and how often the agent actually used it. Neither frequent use nor rare use establishes effectiveness.

Compare failures with successful or otherwise informative episodes facing similar conditions. Include counterexamples that weaken the proposed explanation. Do not select only traces that support your preferred design.

2. Diagnose the limitation of the current harness

For each important pattern, distinguish:

- Observation: what the recorded trajectory directly establishes.
- Mechanism hypothesis: how the harness may have contributed to the behavior.
- Alternative explanations: other causes consistent with the evidence.
- Uncertainty: information that is missing or ambiguous.

Determine whether the problem concerns:

- A capability the agent does not have.
- An existing capability whose interface is difficult to use.
- Observations that are incomplete, excessive, ambiguous, or stale.
- A capability that is available but selected at the wrong time or not used.
- Feedback that is misunderstood or does not support the next decision.
- Coordination between operations, state changes, and limited resources.

These are diagnostic possibilities, not required additions.

Identify which part of the current implementation supports the diagnosis. Do not describe an existing capability as absent merely because it was ineffective in some trajectories.

Separate task failure from exclusion from training. Where metadata permits, identify whether training exclusion follows from invalid or incomplete interaction, postprocessing constraints, group-level reward equality, or another recorded condition. Do not infer the reason from the final reward alone.

Prioritize limitations using recurrence, consequence, avoidable interaction cost, evidence strength, and whether the harness can plausibly affect the behavior. A difficult task or a reasoning error does not automatically imply a missing tool.

When policy versions differ, revisit earlier diagnoses using the current policy–harness trajectories. Treat earlier findings as hypotheses to reconsider, not as established limitations of the current system.
\end{lstlisting}
\end{promptbox}
\caption{Teacher instruction for tool and hook proposals and implementation. Part 2 of 5.}
\label{fig:prompt-harness-2}
\end{figure*}

\begin{figure*}[p]
\centering
\captionsetup{position=bottom,skip=8pt}
\begin{promptbox}{Harness proposal and implementation (3/5)}
\begin{lstlisting}[style=promptcode,firstline=76,lastline=104]
3. Select the appropriate form of improvement

Choose one supported limitation and propose a small, coherent improvement.

Classify every change explicitly:

- Tool change: adds or changes a callable operation, its inputs, action semantics, or returned observations.
- Hook change: adds or changes an intervention triggered by an event and possibly conditioned on retained state.
- Guidance change: changes instructions about selecting, sequencing, or interpreting capabilities.
- Other harness change: specify the affected representation or control mechanism precisely.

Mark each component as added, modified, replaced, or reused.

Distinguish an instruction to perform an action from code that automatically performs or enforces that action. Do not describe guidance as a new tool, or advisory feedback as a mandatory gate.

Consider whether an existing capability can be clarified or adapted before adding another one. Choose the implementation layer that addresses the observed limitation.

A proposal may contain multiple cooperating changes when necessary. For each component, explain:

- The observed problem it addresses.
- What it changes at the relevant decision point.
- Which other components it depends on.
- Why the proposed combination is needed.
- What unnecessary changes have been excluded.

Prefer the smallest coherent mechanism, rather than an arbitrary number of components. Do not bundle unrelated improvements under a broad rationale.

Treat historical outcomes at the scope actually measured. A rejected bundle does not establish that every component is harmful. An accepted bundle does not isolate the benefit of one component. Do not combine results from different policies, task cohorts, or parent configurations into a cumulative improvement claim.
\end{lstlisting}
\end{promptbox}
\caption{Teacher instruction for tool and hook proposals and implementation. Part 3 of 5.}
\label{fig:prompt-harness-3}
\end{figure*}

\begin{figure*}[p]
\centering
\captionsetup{position=bottom,skip=8pt}
\begin{promptbox}{Harness proposal and implementation (4/5)}
\begin{lstlisting}[style=promptcode,firstline=105,lastline=136]
4. Specify the behavioral mechanism and its costs

Explain the expected chain:

observed situation
→ new capability or information
→ changed agent decision or action
→ possible effect on task success and training eligibility

Identify the intended behavioral change and the observations that would weaken your hypothesis.

Consider whether the agent can discover and use the mechanism with the information available at that point. Account for argument preparation, interpretation of feedback, and any extra decisions imposed by the interface.

Describe plausible costs and adverse effects, including additional calls, tokens, execution time, state complexity, interruption, misleading heuristics, or disruption of useful behavior.

Specify when the mechanism should remain inactive, stop repeating, or defer to the agent. Explain how it preserves useful behavior demonstrated by the counterexamples.

If the design produces evidence, state exactly what that evidence establishes. Distinguish an attempted operation from a completed one, structural validity from task correctness, partial checks from comprehensive coverage, and cached observations from fresh execution.

Explain the expected effect on each objective separately. Do not use shorter trajectories, more tool calls, more checks, or more submissions as substitutes for pass rate and trainable rollout ratio.

5. Implement the candidate

Produce concrete code or exact edits using the harness's existing extension points and conventions.

For each tool change, specify:

- Inputs and constraints.
- Actual operations and side effects.
- Returned observations and their meaning.
- Failure behavior and resource bounds.
\end{lstlisting}
\end{promptbox}
\caption{Teacher instruction for tool and hook proposals and implementation. Part 4 of 5.}
\label{fig:prompt-harness-4}
\end{figure*}

\begin{figure*}[p]
\centering
\captionsetup{position=bottom,skip=8pt}
\begin{promptbox}{Harness proposal and implementation (5/5)}
\begin{lstlisting}[style=promptcode,firstline=137,lastline=163]
For each hook change, specify:

- Triggering events and conditions.
- State that is read or retained.
- Feedback or automatic action.
- Repetition limits and reset conditions.
- Conditions under which it remains silent.

For each guidance change, provide the exact instruction text and its integration point. Explain which existing or modified capability it refers to.

Keep the design general across tasks. Do not encode task identifiers, reference solutions, expected answers, hard-coded project paths, or exceptions tailored to individual examples. Use trajectories as evidence for the design, not as solutions to embed in it.

Preserve the task's permissions, evaluator semantics, and training eligibility rules. Make no claims that an action was performed, a test passed, or a behavior improved unless the available evidence establishes it.

Response structure

A. Evidence and diagnosis
Present the prioritized limitation, concrete trajectory and event references, recurrence with denominators where available, successful counterexamples, and alternative explanations.

B. Proposed mechanism
Explain the behavioral change, component dependencies, expected effect on each objective, costs, and uncertainty. Clearly distinguish tools, hooks, guidance, and reused capabilities.

C. Concrete implementation
Provide the code or exact edits, interfaces, trigger and state logic, and instruction text needed to realize the candidate.

D. Predicted behavior
Walk through the relevant decision points in the cited trajectories and explain how the candidate would respond. Include a case where it should not intervene and observations that would weaken the hypothesis. Label these as predictions rather than measured improvements.
\end{lstlisting}
\promptsep
\promptpart{Input template:}
\begin{lstlisting}[style=promptcode]
<round_context>{{ROUND_CONTEXT}}</round_context>
<current_harness>{{CURRENT_HARNESS}}</current_harness>
<runtime_contract>{{RUNTIME_CONTRACT}}</runtime_contract>
<comparison_settings>{{COMPARISON_SETTINGS}}</comparison_settings>
<training_eligibility_rules>{{TRAINING_ELIGIBILITY_RULES}}</training_eligibility_rules>
<preparation_trajectories>{{PREPARATION_TRAJECTORIES}}</preparation_trajectories>
<rollout_metadata>{{ROLLOUT_METADATA}}</rollout_metadata>
<previous_proposals>{{PREVIOUS_PROPOSALS}}</previous_proposals>
\end{lstlisting}
\end{promptbox}
\caption{Teacher instruction for tool and hook proposals and implementation. Part 5 of 5. The input template follows the instruction.}
\label{fig:prompt-harness-5}
\end{figure*}

%% file: acl_latex.bbl
\begin{thebibliography}{41}
\providecommand{\natexlab}[1]{#1}
\providecommand{\url}[1]{\texttt{#1}}
\expandafter\ifx\csname urlstyle\endcsname\relax
  \providecommand{\doi}[1]{doi: #1}\else
  \providecommand{\doi}{doi: \begingroup \urlstyle{rm}\Url}\fi

\bibitem[Jain et~al.(2025)Jain, Singh, Shetty, Zhang, Zheng, Sen, and Stoica]{jain2025r2e}
Naman Jain, Jaskirat Singh, Manish Shetty, Tianjun Zhang, Liang Zheng, Koushik Sen, and Ion Stoica.
\newblock {R2E-Gym}: Procedural environment generation and hybrid verifiers for scaling open-weights {SWE} agents.
\newblock In \emph{Second Conference on Language Modeling}, 2025.
\newblock URL \url{https://openreview.net/forum?id=7evvwwdo3z}.

\bibitem[Yang et~al.(2025{\natexlab{a}})Yang, Lieret, Jimenez, Wettig, Khandpur, Zhang, Hui, Press, Schmidt, and Yang]{yang2025swesmith}
John Yang, Kilian Lieret, Carlos~E. Jimenez, Alexander Wettig, Kabir Khandpur, Yanzhe Zhang, Binyuan Hui, Ofir Press, Ludwig Schmidt, and Diyi Yang.
\newblock {SWE-smith}: Scaling data for software engineering agents.
\newblock In \emph{Advances in Neural Information Processing Systems}, 2025{\natexlab{a}}.
\newblock URL \url{https://proceedings.neurips.cc/paper_files/paper/2025/hash/8b86cf5ace600c48fd188efbb8dedec8-Abstract-Datasets_and_Benchmarks_Track.html}.

\bibitem[Xi et~al.(2026)Xi, Huang, Liao, Huang, Liu, Guo, Yang, Zheng, Ye, Zhang, Chen, He, Ding, Li, Chen, Du, Yao, Xu, Chen, Gui, Wu, Zhang, Huang, and Jiang]{xi2026agentgymrl}
Zhiheng Xi, Jixuan Huang, Chenyang Liao, Baodai Huang, Jiaqi Liu, Honglin Guo, Yajie Yang, Rui Zheng, Junjie Ye, Jiazheng Zhang, Wenxiang Chen, Wei He, Yiwen Ding, Guanyu Li, Zehui Chen, Zhengyin Du, Xuesong Yao, Yufei Xu, Jiecao Chen, Tao Gui, Zuxuan Wu, Qi~Zhang, Xuanjing Huang, and Yu-Gang Jiang.
\newblock {AgentGym-RL}: An open-source framework to train {LLM} agents for long-horizon decision making via multi-turn {RL}.
\newblock In \emph{International Conference on Learning Representations}, 2026.
\newblock URL \url{https://openreview.net/forum?id=ZgCCDwcGwn}.

\bibitem[Chen et~al.(2026)Chen, Xiao, Zhu, Yuan, Zhang, and Wang]{chen2026coharness}
Zhengyu Chen, Teng Xiao, Huaisheng Zhu, Yige Yuan, Luan Zhang, and Jingang Wang.
\newblock {Co-Harness}: Co-evolving harnesses and model weights for {LLM} agents.
\newblock \emph{arXiv preprint arXiv:2607.22688}, 2026.
\newblock URL \url{https://arxiv.org/abs/2607.22688}.

\bibitem[Lee et~al.(2026{\natexlab{a}})Lee, Nair, Zhang, Lee, Khattab, and Finn]{lee2026metaharness}
Yoonho Lee, Roshen Nair, Qizheng Zhang, Kangwook Lee, Omar Khattab, and Chelsea Finn.
\newblock {Meta-Harness}: End-to-end optimization of model harnesses.
\newblock \emph{arXiv preprint arXiv:2603.28052}, 2026{\natexlab{a}}.
\newblock URL \url{https://arxiv.org/abs/2603.28052}.

\bibitem[Jiang et~al.(2026)Jiang, Chu, Tian, Zhang, Yang, Yang, Liu, Lv, and Li]{jiang2026harnessevolve}
Wen Jiang, Mingmin Chu, Yimeng Tian, Qianxin Zhang, Haofei Yang, Rui Yang, Yang Liu, Tao Lv, and Fangming Li.
\newblock {HarnessEvolve}: Learning from reference trajectories for reliable agent self-evolution.
\newblock \emph{arXiv preprint arXiv:2609.00829}, 2026.
\newblock URL \url{https://arxiv.org/abs/2609.00829}.

\bibitem[Du et~al.(2026)Du, Wang, Xu, Xu, Tan, Zhao, Yang, Xu, Kong, Wei, Liu, and Zhu]{du2026livingharness}
Yuetian Du, Yucheng Wang, He~Xu, Jiexu Xu, Shanwen Tan, Bing Zhao, Boyu Yang, Zhijie Xu, Ming Kong, Hu~Wei, Jie Liu, and Qiang Zhu.
\newblock {Living-Harness} is an interactive-agent evolver.
\newblock \emph{arXiv preprint arXiv:2607.26598}, 2026.
\newblock URL \url{https://arxiv.org/abs/2607.26598}.

\bibitem[Xiao et~al.(2026)Xiao, Jiao, Wang, Wang, Zhao, Wei, Zhang, and Qu]{xiao2026socraticswe}
Chuan Xiao, Zhengbo Jiao, Shaobo Wang, Wei Wang, Bing Zhao, Hu~Wei, Linfeng Zhang, and Lin Qu.
\newblock {Socratic-SWE}: Self-evolving coding agents via trace-derived agent skills.
\newblock \emph{arXiv preprint arXiv:2606.07412}, 2026.
\newblock URL \url{https://arxiv.org/abs/2606.07412}.

\bibitem[Wang et~al.(2026)Wang, Li, Yang, and Gao]{hhd2026}
Shengjie Wang, Guanghe Li, Zonghan Yang, and Yang Gao.
\newblock Hindsight hint distillation: Scaffolded reasoning for {SWE} agents from {CoT}-free answers.
\newblock \emph{arXiv preprint arXiv:2605.11556}, 2026.
\newblock URL \url{https://arxiv.org/abs/2605.11556}.

\bibitem[Shi et~al.(2026)Shi, Ma, Wang, Tan, Li, Chen, and Zhu]{pats2026}
Yipeng Shi, Zhipeng Ma, Yue Wang, Qitai Tan, Yang Li, Peng Chen, and Zhengzhou Zhu.
\newblock {PATS}: Policy-aware training scaffolding for agentic reinforcement learning.
\newblock \emph{arXiv preprint arXiv:2607.21419}, 2026.
\newblock URL \url{https://arxiv.org/abs/2607.21419}.

\bibitem[Shao et~al.(2024)Shao, Wang, Zhu, Xu, Song, Bi, Zhang, Zhang, Li, Wu, and Guo]{shao2024deepseekmath}
Zhihong Shao, Peiyi Wang, Qihao Zhu, Runxin Xu, Junxiao Song, Xiao Bi, Haowei Zhang, Mingchuan Zhang, Y.~K. Li, Y.~Wu, and Daya Guo.
\newblock {DeepSeekMath}: Pushing the limits of mathematical reasoning in open language models.
\newblock \emph{arXiv preprint arXiv:2402.03300}, 2024.

\bibitem[Chowdhury et~al.(2024)Chowdhury, Aung, Shern, Jaffe, Sherburn, Starace, Mays, Dias, Aljubeh, Glaese, Jimenez, Yang, Ho, Patwardhan, Liu, and Madry]{chowdhury2024verified}
Neil Chowdhury, James Aung, Chan~Jun Shern, Oliver Jaffe, Dane Sherburn, Giulio Starace, Evan Mays, Rachel Dias, Marwan Aljubeh, Mia Glaese, Carlos~E. Jimenez, John Yang, Leyton Ho, Tejal Patwardhan, Kevin Liu, and Aleksander Madry.
\newblock Introducing {SWE}-bench verified.
\newblock OpenAI blog, 2024.
\newblock URL \url{https://openai.com/index/introducing-swe-bench-verified/}.

\bibitem[Deng et~al.(2025)Deng, Da, Pan, He, Ide, Garg, Lauffer, Park, Pasari, Rane, Sampath, Krishnan, Kundurthy, Hendryx, Wang, Bharadwaj, Holm, Aluri, Zhang, Jacobson, Liu, and Kenstler]{deng2025swebenchpro}
Xiang Deng, Jeff Da, Edwin Pan, Yannis~Yiming He, Charles Ide, Kanak Garg, Niklas Lauffer, Andrew Park, Nitin Pasari, Chetan Rane, Karmini Sampath, Maya Krishnan, Srivatsa Kundurthy, Sean Hendryx, Zifan Wang, Vijay Bharadwaj, Jeff Holm, Raja Aluri, Chen Bo~Calvin Zhang, Noah Jacobson, Bing Liu, and Brad Kenstler.
\newblock {SWE-Bench Pro}: Can ai agents solve long-horizon software engineering tasks?
\newblock \emph{arXiv preprint arXiv:2509.16941}, 2025.
\newblock URL \url{https://arxiv.org/abs/2509.16941}.

\bibitem[Yao et~al.(2023)Yao, Zhao, Yu, Du, Shafran, Narasimhan, and Cao]{yao2023react}
Shunyu Yao, Jeffrey Zhao, Dian Yu, Nan Du, Izhak Shafran, Karthik Narasimhan, and Yuan Cao.
\newblock {ReAct}: Synergizing reasoning and acting in language models.
\newblock In \emph{International Conference on Learning Representations}, 2023.
\newblock URL \url{https://arxiv.org/abs/2210.03629}.

\bibitem[Yang et~al.(2024)Yang, Jimenez, Wettig, Lieret, Yao, Narasimhan, and Press]{yang2024sweagent}
John Yang, Carlos~E. Jimenez, Alexander Wettig, Kilian Lieret, Shunyu Yao, Karthik Narasimhan, and Ofir Press.
\newblock {SWE}-agent: Agent-computer interfaces enable automated software engineering.
\newblock In \emph{Advances in Neural Information Processing Systems}, 2024.
\newblock URL \url{https://arxiv.org/abs/2405.15793}.

\bibitem[Han et~al.(2026)Han, Zhang, Song, Fang, Chen, Sun, and Hu]{han2026sweskillsbench}
Tingxu Han, Yi~Zhang, Wei Song, Chunrong Fang, Zhenyu Chen, Youcheng Sun, and Lijie Hu.
\newblock {SWE-Skills-Bench}: Do agent skills actually help in real-world software engineering?
\newblock \emph{arXiv preprint arXiv:2603.15401}, 2026.
\newblock URL \url{https://arxiv.org/abs/2603.15401}.

\bibitem[Suzgun et~al.(2025)Suzgun, Yuksekgonul, Bianchi, Jurafsky, and Zou]{suzgun2025dynamiccheatsheet}
Mirac Suzgun, Mert Yuksekgonul, Federico Bianchi, Dan Jurafsky, and James Zou.
\newblock Dynamic cheatsheet: Test-time learning with adaptive memory.
\newblock \emph{arXiv preprint arXiv:2504.07952}, 2025.
\newblock URL \url{https://arxiv.org/abs/2504.07952}.

\bibitem[Xia et~al.(2026)Xia, Chen, Wang, Liu, Zeng, Wang, Han, Zhou, Zhao, Chen, Zheng, Xie, and Yao]{xia2026skillrl}
Peng Xia, Jianwen Chen, Hanyang Wang, Jiaqi Liu, Kaide Zeng, Yu~Wang, Siwei Han, Yiyang Zhou, Xujiang Zhao, Haifeng Chen, Zeyu Zheng, Cihang Xie, and Huaxiu Yao.
\newblock {SkillRL}: Evolving agents via recursive skill-augmented reinforcement learning.
\newblock \emph{arXiv preprint arXiv:2602.08234}, 2026.
\newblock URL \url{https://arxiv.org/abs/2602.08234}.

\bibitem[Yu et~al.(2025)Yu, Zhang, Zhu, Yuan, Zuo, Yue, Dai, Fan, Liu, Liu, Liu, Liu, Lin, Lin, Ma, Sheng, Tong, Zhang, Zhang, Zhang, Zhang, Zhu, Zhu, Chen, Chen, Wang, Yu, Song, Wei, Zhou, Liu, Ma, Zhang, Yan, Wu, and Wang]{yu2025dapo}
Qiying Yu, Zheng Zhang, Ruofei Zhu, Yufeng Yuan, Xiaochen Zuo, Yu~Yue, Weinan Dai, Tiantian Fan, Gaohong Liu, Juncai Liu, Lingjun Liu, Xin Liu, Haibin Lin, Zhiqi Lin, Bole Ma, Guangming Sheng, Yuxuan Tong, Chi Zhang, Mofan Zhang, Ru~Zhang, Wang Zhang, Hang Zhu, Jinhua Zhu, Jiaze Chen, Jiangjie Chen, Chengyi Wang, Hongli Yu, Yuxuan Song, Xiangpeng Wei, Hao Zhou, Jingjing Liu, Wei-Ying Ma, Ya-Qin Zhang, Lin Yan, Yonghui Wu, and Mingxuan Wang.
\newblock {DAPO}: An open-source {LLM} reinforcement learning system at scale.
\newblock In \emph{Advances in Neural Information Processing Systems}, 2025.
\newblock URL \url{https://proceedings.neurips.cc/paper_files/paper/2025/hash/a4277440d50f1f15d2cb4c14f7e0c0d2-Abstract-Conference.html}.

\bibitem[{Qwen Team}(2026)]{qwen2026qwen35}
{Qwen Team}.
\newblock {Qwen3.5}: Towards native multimodal agents, February 2026.
\newblock URL \url{https://qwen.ai/blog?id=qwen3.5}.

\bibitem[{Z.ai}(2026)]{zai2026glm53flash}
{Z.ai}.
\newblock {GLM-5.3-Flash/FlashX}: Model overview and api access.
\newblock Z.AI Developer Documentation, 2026.
\newblock URL \url{https://docs.z.ai/guides/vlm/glm-5.3-flash}.
\newblock Accessed October 20, 2026.

\bibitem[{OpenAI}(2026)]{openai2026gpt56sol}
{OpenAI}.
\newblock {GPT-5.6 Sol} model.
\newblock OpenAI API documentation, 2026.
\newblock URL \url{https://developers.openai.com/api/docs/models/gpt-5.6-sol}.
\newblock Accessed October 20, 2026.

\bibitem[{OpenAI}(2025)]{openai2025codex}
{OpenAI}.
\newblock {Codex} is now generally available.
\newblock OpenAI, October 2025.
\newblock URL \url{https://openai.com/index/codex-now-generally-available/}.

\bibitem[Wang et~al.(2025)Wang, Li, Song, Xu, Tang, Zhuge, Pan, Song, Li, Singh, Tran, Li, Ma, Zheng, Qian, Shao, Muennighoff, Zhang, Hui, Lin, Brennan, Peng, Ji, and Neubig]{wang2025openhands}
Xingyao Wang, Boxuan Li, Yufan Song, Frank~F. Xu, Xiangru Tang, Mingchen Zhuge, Jiayi Pan, Yueqi Song, Bowen Li, Jaskirat Singh, Hoang~H. Tran, Fuqiang Li, Ren Ma, Mingzhang Zheng, Bill Qian, Yanjun Shao, Niklas Muennighoff, Yizhe Zhang, Binyuan Hui, Junyang Lin, Robert Brennan, Hao Peng, Heng Ji, and Graham Neubig.
\newblock {OpenHands}: An open platform for {AI} software developers as generalist agents.
\newblock In \emph{International Conference on Learning Representations}, 2025.
\newblock URL \url{https://arxiv.org/abs/2407.16741}.

\bibitem[Jimenez et~al.(2024)Jimenez, Yang, Wettig, Yao, Pei, Press, and Narasimhan]{jimenez2024swebench}
Carlos~E. Jimenez, John Yang, Alexander Wettig, Shunyu Yao, Kexin Pei, Ofir Press, and Karthik Narasimhan.
\newblock {SWE}-bench: Can language models resolve real-world {GitHub} issues?
\newblock In \emph{International Conference on Learning Representations}, 2024.

\bibitem[Wang et~al.(2024)Wang, Xie, Jiang, Mandlekar, Xiao, Zhu, Fan, and Anandkumar]{wang2024voyager}
Guanzhi Wang, Yuqi Xie, Yunfan Jiang, Ajay Mandlekar, Chaowei Xiao, Yuke Zhu, Linxi Fan, and Anima Anandkumar.
\newblock {Voyager}: An open-ended embodied agent with large language models.
\newblock \emph{Transactions on Machine Learning Research}, 2024.
\newblock URL \url{https://openreview.net/forum?id=ehfRiF0R3a}.

\bibitem[Zhang et~al.(2026)Zhang, Hu, Upasani, Ma, Hong, Kamanuru, Rainton, Wu, Ji, Li, Thakker, Zou, and Olukotun]{zhang2026agentic}
Qizheng Zhang, Changran Hu, Shubhangi Upasani, Boyuan Ma, Fenglu Hong, Vamsidhar Kamanuru, Jay Rainton, Chen Wu, Mengmeng Ji, Hanchen Li, Urmish Thakker, James Zou, and Kunle Olukotun.
\newblock Agentic context engineering: Evolving contexts for self-improving language models.
\newblock In \emph{International Conference on Learning Representations}, 2026.
\newblock URL \url{https://arxiv.org/abs/2510.04618}.

\bibitem[Shinn et~al.(2023)Shinn, Cassano, Gopinath, Narasimhan, and Yao]{shinn2023reflexion}
Noah Shinn, Federico Cassano, Ashwin Gopinath, Karthik Narasimhan, and Shunyu Yao.
\newblock {Reflexion}: Language agents with verbal reinforcement learning.
\newblock In \emph{Advances in Neural Information Processing Systems}, 2023.
\newblock URL \url{https://proceedings.neurips.cc/paper_files/paper/2023/hash/1b44b878bb782e6954cd888628510e90-Abstract-Conference.html}.

\bibitem[Madaan et~al.(2023)Madaan, Tandon, Gupta, Hallinan, Gao, Wiegreffe, Alon, Dziri, Prabhumoye, Yang, Gupta, Majumder, Hermann, Welleck, Yazdanbakhsh, and Clark]{madaan2023selfrefine}
Aman Madaan, Niket Tandon, Prakhar Gupta, Skyler Hallinan, Luyu Gao, Sarah Wiegreffe, Uri Alon, Nouha Dziri, Shrimai Prabhumoye, Yiming Yang, Shashank Gupta, Bodhisattwa~Prasad Majumder, Katherine Hermann, Sean Welleck, Amir Yazdanbakhsh, and Peter Clark.
\newblock {Self-Refine}: Iterative refinement with self-feedback.
\newblock In \emph{Advances in Neural Information Processing Systems}, 2023.
\newblock URL \url{https://proceedings.neurips.cc/paper_files/paper/2023/hash/91edff07232fb1b55a505a9e9f6c0ff3-Abstract-Conference.html}.

\bibitem[Pan et~al.(2026)Pan, Liu, Lin, Zeng, Tang, Zhou, Lu, and Jia]{pan2026evolving}
Wenbo Pan, Shujie Liu, Chin-Yew Lin, Jingying Zeng, Xianfeng Tang, Xiangyang Zhou, Yan Lu, and Xiaohua Jia.
\newblock Evolving agents in the dark: Retrospective harness optimization via self-preference.
\newblock \emph{arXiv preprint arXiv:2606.05922}, 2026.

\bibitem[Lee et~al.(2026{\natexlab{b}})Lee, Xu, Seely, Lee, Zaharia, and Tang]{lee2026rhi}
Hyunin Lee, Jinglue Xu, Jeffrey Seely, Donghyun Lee, Matei Zaharia, and Yujin Tang.
\newblock Recursive harness self-improvement.
\newblock \emph{arXiv preprint arXiv:2607.15524}, 2026{\natexlab{b}}.
\newblock URL \url{https://arxiv.org/abs/2607.15524}.

\bibitem[Yang et~al.(2026)Yang, Ma, Huang, Hu, Wang, and Chu]{yang2026coevolve}
Shidong Yang, Ziyu Ma, Tongwen Huang, Yiming Hu, Yong Wang, and Xiangxiang Chu.
\newblock {CoEvolve}: Training {LLM} agents via agent-data mutual evolution.
\newblock In \emph{Proceedings of the 64th Annual Meeting of the Association for Computational Linguistics (Volume 1: Long Papers)}, pages 23015--23036. Association for Computational Linguistics, 2026.
\newblock URL \url{https://aclanthology.org/2026.acl-long.1055/}.

\bibitem[Feng et~al.(2026)Feng, Song, Li, Liu, and Shao]{feng2026searl}
Xinshun Feng, Xinhao Song, Lijun Li, Gongshen Liu, and Jing Shao.
\newblock {SEARL}: Joint optimization of policy and tool graph memory for self-evolving agents.
\newblock In \emph{Proceedings of the 64th Annual Meeting of the Association for Computational Linguistics (Volume 1: Long Papers)}, pages 24518--24535. Association for Computational Linguistics, 2026.
\newblock URL \url{https://aclanthology.org/2026.acl-long.1125/}.

\bibitem[Song et~al.(2026)Song, Bai, Yang, Li, Deng, He, Zhang, Cheng, Xing, Yun, Chen, Li, Chang, Hao, Tao, Yang, Dai, Zhao, Tang, and Wen]{clawgym2026}
Huatong Song, Fei Bai, Ming Yang, Renyuan Li, Jia Deng, Jujie He, Zhange Zhang, Daixuan Cheng, Yan Xing, Qi~Yun, Xuxing Chen, Danyang Li, Feng Chang, Chuan Hao, Ran Tao, Jian Yang, Bryan Dai, Wayne~Xin Zhao, Mingjie Tang, and Ji-Rong Wen.
\newblock {ClawGym II}: Exploring black-box {RL} on agent harness.
\newblock \emph{arXiv preprint arXiv:2608.16798}, 2026.
\newblock URL \url{https://arxiv.org/abs/2608.16798}.

\bibitem[Yang et~al.(2025{\natexlab{b}})Yang, Wang, Fu, He, Xiong, Liu, Miao, Gao, Wang, Ma, Li, Liu, Hu, Zhang, Wang, Chen, Sung, Liu, Gao, Yang, and Liu]{yang2025kimidev}
Zonghan Yang, Shengjie Wang, Kelin Fu, Wenyang He, Weimin Xiong, Yibo Liu, Yibo Miao, Bofei Gao, Yejie Wang, Yingwei Ma, Yanhao Li, Yue Liu, Zhenxing Hu, Kaitai Zhang, Shuyi Wang, Huarong Chen, Flood Sung, Yang Liu, Yang Gao, Zhilin Yang, and Tianyu Liu.
\newblock Kimi-dev: Agentless training as skill prior for {SWE}-agents.
\newblock \emph{arXiv preprint arXiv:2509.23045}, 2025{\natexlab{b}}.
\newblock URL \url{https://arxiv.org/abs/2509.23045}.

\bibitem[Cao et~al.(2025)Cao, Li, Zhao, Yuan, Hegde, Chen, Ruan, Griggs, Liu, Tang, Liaw, Moritz, Zaharia, Gonzalez, and Stoica]{cao2025skyrlagent}
Shiyi Cao, Dacheng Li, Fangzhou Zhao, Shuo Yuan, Sumanth~R. Hegde, Connor Chen, Charlie Ruan, Tyler Griggs, Shu Liu, Eric Tang, Richard Liaw, Philipp Moritz, Matei Zaharia, Joseph~E. Gonzalez, and Ion Stoica.
\newblock {SkyRL-Agent}: Efficient {RL} training for multi-turn {LLM} agent.
\newblock \emph{arXiv preprint arXiv:2511.16108}, 2025.
\newblock URL \url{https://arxiv.org/abs/2511.16108}.

\bibitem[Kim et~al.(2026)Kim, Cho, Kwak, Kwon, Wang, Yang, Zhang, Wei, and Yeo]{horizon2026}
Sunghwan Kim, Junhee Cho, Beong-woo Kwak, Taeyoon Kwon, Liang Wang, Nan Yang, Xingxing Zhang, Furu Wei, and Jinyoung Yeo.
\newblock On training large language models for long-horizon tasks: An empirical study of horizon length.
\newblock \emph{arXiv preprint arXiv:2605.02572}, 2026.
\newblock URL \url{https://arxiv.org/html/2605.02572v1}.

\bibitem[Zhu et~al.(2025)Zhu, Xie, Lv, and {slime Contributors}]{slime2026}
Zilin Zhu, Chengxing Xie, Xin Lv, and {slime Contributors}.
\newblock slime: An {LLM} post-training framework for {RL} scaling.
\newblock GitHub repository, 2025.
\newblock URL \url{https://github.com/THUDM/slime}.
\newblock Accessed September 18, 2026.

\bibitem[{OpenCode Contributors}(2026)]{opencode2026}
{OpenCode Contributors}.
\newblock {OpenCode}: The open source coding agent.
\newblock GitHub repository, 2026.
\newblock URL \url{https://github.com/anomalyco/opencode}.
\newblock Accessed September 18, 2026.

\bibitem[Luo et~al.(2025)Luo, Jain, Singh, Tan, Patel, Wu, Ariyak, Cai, Venkat, Zhu, Athiwaratkun, Roongta, Zhang, Li, Popa, Sen, and Stoica]{luo2025deepswe}
Michael Luo, Naman Jain, Jaskirat Singh, Sijun Tan, Ameen Patel, Qingyang Wu, Alpay Ariyak, Colin Cai, Tarun Venkat, Shang Zhu, Ben Athiwaratkun, Manan Roongta, Ce~Zhang, Li~Erran Li, Raluca~Ada Popa, Koushik Sen, and Ion Stoica.
\newblock {DeepSWE}: Training a fully open-sourced, state-of-the-art coding agent by scaling {RL}.
\newblock Together AI blog, 2025.
\newblock URL \url{https://www.together.ai/blog/deepswe}.

\bibitem[Miller(2024)]{miller2024errorbars}
Evan Miller.
\newblock Adding error bars to evals: A statistical approach to language model evaluations.
\newblock \emph{arXiv preprint arXiv:2411.00640}, 2024.
\newblock URL \url{https://arxiv.org/abs/2411.00640}.

\end{thebibliography}
